\documentstyle[12pt,epsfig]{article}
\newcommand\pubnumber{BNL--67545 \\ 
        FERMILAB--Pub--00/152 \\ 
         LBNL--46299 \\ 
           SLAC-PUB-8495\\ UCRL--ID--139524}               
\newcommand\pubdate{July, 2000}

\def\Title#1{\begin{center} {\Large #1 } \end{center}}
\def\Author#1{\begin{center}{ \sc #1} \end{center}}

\newcommand\pubblock{\rightline{\begin{tabular}{l} \pubnumber\\
         \pubdate  \end{tabular}}}
\newenvironment{Abstract}{\begin{quotation} \begin{center}
                       ABSTRACT
     \end{center}\bigskip  }{\end{quotation}}
\def\beq{\begin{equation}}
\def\eeq#1{\label{#1}\end{equation}}
\def\eeqn{\end{equation}}
\def\beqa{\begin{eqnarray}}
\def\eeqa#1{\label{#1}\end{eqnarray}}
\def\eeqan{\end{eqnarray}}

\def\leqn#1{(\ref{#1})}
\def\Acknowledgements{\bigskip  \bigskip \begin{center} \begin{large}
             \bf ACKNOWLEDGEMENTS \end{large}\end{center}}

\let\bar=\overbar
\def\Dslash{\not{\hbox{\kern-4pt $D$}}}
\def\dslash{\not{\hbox{\kern-2pt $\del$}}}
\def\half{\frac{1}{2}}
\def\thalf{\frac{3}{2}}

\def\L{{\cal L}}

\def\O{{\cal O}}

\def\del{\partial}

\def\VEV#1{\left\langle{ #1} \right\rangle}

\def\ee{e^+e^-}
\def\sstw{\sin^2\theta_w}
\def\cstw{\cos^2\theta_w}
\def\mz{m_Z}

\def\mw{m_W}

\def\mh{m_h}

\def\msb{{\bar{\scriptsize M \kern -1pt S}}}

\def\ch#1{\widetilde\chi^+_{#1}}
\def\chm#1{\widetilde\chi^-_{#1}}
\def\neu#1{\widetilde\chi^0_{#1}}
\def\s#1{\widetilde{#1}}

\def\etal{{\it et al.}}

\def\eg{{\it e.g.}}
\def\lunit{cm$^{-2}$sec$^{-1}$}

\newcommand{\lsim}{\mathrel{\lower4pt\hbox{$\sim$}}
\hskip-14.5pt\raise1.6pt\hbox{$<$}\;}

\newcommand{\gsim}{\mathrel{\lower4pt\hbox{$\sim$}}
\hskip-13.5pt\raise1.6pt\hbox{$>$}\;}

\begin{document}
\begin{titlepage}
\pubblock

\vfill
\Title{The Case for a 500 GeV $\ee$ Linear Collider}
\vfill
\Author{American Linear Collider Working Group}
\medskip
\begin{center}
J.~Bagger$^{11*}$,            
C.~Baltay$^{28*}$,           
T.~Barker$^{7}$,
T.~Barklow$^{26}$,            
U.~Baur$^{18}$,
T.~Bolton$^{12}$,
J.~Brau$^{20}$,
M.~Breidenbach$^{26}$,
D.~Burke$^{26}$,
P.~Burrows$^{21}$,
L.~Dixon$^{26}$,
H.~E.~Fisk$^{8}$,
R.~Frey$^{20}$,                
D.~Gerdes$^{17}$,
N.~Graf$^{26}$,
P.~D.~Grannis$^{19*}$,         
H.~E.~Haber$^{4}$,               
C.~Hearty$^{1}$,
S.~Hertzbach$^{15}$,
C.~Heusch$^{4}$,
J.~Hewett$^{26}$,
R.~Hollebeek$^{22}$,
R.~Jacobsen$^{13}$,
J.~Jaros$^{26}$,
T.~Kamon$^{27}$,               
D.~Karlen$^{6}$,               
D.~Koltick$^{23}$,
A.~Kronfeld$^{8}$,
W.~Marciano$^{2}$,             
T.~Markiewicz$^{26}$,
H.~Murayama$^{13}$,
U.~Nauenberg$^{7}$,
L.~Orr$^{24}$,               
F.~Paige$^{2}$,              
A.~Para$^{8}$,
M.~E.~Peskin$^{26*}$,          
F.~Porter$^{5}$,
K.~Riles$^{17}$,            
M.~Ronan$^{13}$,
L.~Rosenberg$^{16}$,
B.~Schumm$^{4}$,
R.~Stroynowski$^{25}$,
S.~Tkaczyk$^{8}$,             
A.~S.~Turcot$^{2}$,            
K.~van~Bibber$^{14}$,         
R.~van~Kooten$^{10}$,
J.~D.~Wells$^{3}$,              
H.~Yamamoto$^{9}$
\end{center}

\vfill
\begin{Abstract}
Several proposals are  being developed around the world
for an $\ee$ linear collider with an initial center of mass energy of 
500 GeV.  In this paper, we will discuss why a project of this type
deserves priority as the next major initiative in high energy physics.
\end{Abstract}
\vfill

\begin{small} \noindent
Work supported in part by the US Department of Energy under
contracts DE--AC02--76CH03000,
DE--AC02--98CH10886, DE--AC03--76SF00098, DE--AC03--76SF00515, and
W--7405--ENG--048.
\end{small}
\newpage

\centerline{ $^{*}$ editorial board for this paper}
\centerline{$^{1}$ University of British Columbia, Vancouver, 
                          BC V6T 1Z1, CANADA}
\centerline{ $^{2}$ Brookhaven National Laboratory, Upton, NY 11973}
\centerline{ $^{3}$ University of California, Davis, CA 95616}
\centerline{ $^{4}$ University of California, Santa Cruz, CA 95064}
\centerline{ $^{5}$ California Institute of Technology, Pasadena, CA 91125}
\centerline{ $^{6}$ Carleton University, Ottawa, ON K1S 5B6, CANADA}
\centerline{ $^{7}$ University of Colorado, Boulder, CO 80309}
\centerline{ $^{8}$ Fermi National Accelerator Laboratory, Batavia, IL 60510}
\centerline{ $^{9}$ University of Hawaii, Honolulu, HI 96822}
\centerline{ $^{10}$ Indiana University, Bloomington, IN 47405}
\centerline{ $^{11}$ Johns Hopkins University, Baltimore, MD 21218}
\centerline{ $^{12}$ Kansas State University, Manhattan, KS 66506}
\centerline{ $^{13}$ Lawrence Berkeley National Laboratory, Berkeley, CA
           94720}
\centerline{ $^{14}$ Lawrence Livermore National Laboratory, Livermore, CA
        94551}
\centerline{ $^{15}$ University of Massachusetts, Amherst, MA 01003}
\centerline{ $^{16}$ Massachussetts Institute of Technology, 
                                 Cambridge, MA 02139}
\centerline{ $^{17}$ University of Michigan, Ann Arbor MI 48109}
\centerline{ $^{18}$ State University of New York, Buffalo, NY 14260}
\centerline{ $^{19}$ State University of New York, Stony Brook, NY 11794}
\centerline{ $^{20}$ University of Oregon, Eugene, OR 97403}
\centerline{ $^{21}$ Oxford University, Oxford OX1 3RH, UK}
\centerline{ $^{22}$ University of Pennslyvania, Philadelphia, PA 19104}
\centerline{ $^{23}$ Purdue University, West Lafayette, IN 47907}
\centerline{ $^{24}$ University of Rochester, Rochester, NY 14627}
\centerline{ $^{25}$ Southern Methodist University, Dallas, TX 75275}
\centerline{ $^{26}$ Stanford Linear Accelerator Center, Stanford, CA 94309}
\centerline{ $^{27}$ Texas A\&M University, College Station, TX 77843}
\centerline{ $^{28}$ Yale University, New Haven, CT 06520}

\end{titlepage}
\def\thefootnote{\fnsymbol{footnote}}
\setcounter{footnote}{0}

\tableofcontents
\newpage
\section{Introduction}

Those of us who have chosen to work in elementary particle physics
have taken on the task of uncovering the laws of Nature
at the smallest distance scales.
The process is an excavation, and as such, the work proceeds
through various stages.  During the past ten years,
experiments have clarified the basic structure of the
strong, weak, and electromagnetic interactions through measurements
of exquisite precision.  Now the next stage is about to begin.

The structure of the electroweak interactions, confirmed in great
detail by recent experiments, requires a new threshold in
fundamental physics at distances or energies within a factor of
ten beyond those we can currently probe.  More detailed aspects of
the data argue that this threshold is close at hand.  In the next
decade, we will carry out the first experiments that move beyond
this threshold, perhaps at the Fermilab Tevatron, almost certainly
at the CERN LHC.

Many measurements of this new physics will be made at
these hadron colliders.  In this document we will argue
that electron-positron colliders also have
an important role to play.  Because the electron is an
essentially structureless particle which interacts through the
precisely
 calculable
weak and electromagnetic interactions, an
$\ee$ collider can unambiguously determine the spins and quantum
numbers of new particles.  Cross section and branching ratio
measurements are straightforward and can be readily compared to models
for the underlying physics.  Electron beam polarization allows
experiments to distinguish electroweak quantum numbers and measure
important mixing angles.   During the next few years, hadron
colliders will likely discover the agents of electroweak symmetry
breaking.  But electron-positron experiments will also be
necessary to completely determine the properties of the new
particles.

We believe that a number of new developments call for
the start of
construction of a high luminosity 500 GeV $\ee$ collider in this decade.
First, precision measurements from experiments at CERN,
Fermilab and SLAC suggest that important new physics is within
range of this machine.  Second, the necessary technologies
have been developed to the point where it is feasible to construct
the collider.  Third, these technologies, and others still
under development, should allow the collider to be upgraded
to TeV and even multi-TeV energies.  For all of these reasons,
we believe that the time is right to design and construct
a high luminosity 500 GeV $\ee$ linear collider.

In this paper, we  formulate the physics case for this
machine.  The elements of the  argument are:
\begin{enumerate}
\item New physics processes should appear at a 500 GeV collider.
In particular, precision data indicate that the
Higgs boson should be accessible to this machine.  If it is,
the collider will definitively test whether the Higgs boson is
responsible for generating the masses of the quarks, leptons,
and gauge bosons of the Standard Model.
\item There are good reasons to believe that there is other
new physics at the TeV scale.  Across the range of models,
$\ee$ collider experiments add crucial information to that
available from hadron collider experiments.  They will
dramatically clarify our understanding of TeV scale physics.
\item A 500 GeV collider is a critical first step toward a
higher energy $\ee$ collider.  We believe that such a machine
is likely to be needed for the complete elucidation of the next
set of physical laws.
\end{enumerate}

This paper will proceed as follows:  In Section 2, we will
discuss the future of high energy physics from a long-term
perspective.  We will briefly review the recent developments
that have clarified the structure of elementary particle
interactions, the challenges posed by the next scale in physics,
and the need for
higher energy lepton and hadron colliders. In Section 3, we
will briefly describe the current designs of 500 GeV
$\ee$ colliders and the technologies that will enable them
to be upgraded to higher energy.  This discussion will define
the basic accelerator specifications that we will explore
in this study: center of mass energies up to 500 GeV, and
luminosity samples of 200 fb$^{-1}$ to 600 fb$^{-1}$. In
Section 4, we will give the arguments that new physics should
appear at 500 GeV.  In Section 5, we will describe some of
the important measurements that could be made at a 500 GeV
collider,
or with high luminosity measurements at the $Z$ pole or the $WW$  
threshold.
  In Section 6, we will describe additional measurements
for which the required energy is less certain but which, when
they are kinematically accessible in $\ee$ collisions, will
beautifully enhance the results of the LHC. Section 7 contains
our conclusions.

There is an enormous literature on the physics capabilities
of $\ee$ colliders at energies of 500 GeV and above.  Our goal
in this document is to summarize and focus this information.
Much more information about the capabilities of a high energy
$\ee$ linear collider can be found in
\cite{Murayama,Accomando,SnowmassPurple,Sitges} and
references therein.

Before beginning our discussion, we would like to comment on
 three
related issues.  The first is the role of the LHC.  The
ATLAS and CMS experiments at the LHC
are likely to be the most important
high energy physics experiments of the decade, precisely because
they will be the first experiments whose energy is clearly in
the regime of new physics.  The linear collider does not need
to compete directly with the LHC in terms of energy; instead,
its physics program should  complement the LHC by adding important
new information.  It is just for this reason that we must look
at the strengths and weaknesses of the LHC when we build
the case for an $\ee$ linear collider.

The second concerns the competing linear collider
technologies, the approach of NLC and JLC, with warm copper
accelerating structures, and that of TESLA, with superconducting
RF cavities.  From the point of view of the physics, the similarities
of these proposals are more important than their differences.
Both schemes  are capable of high luminosity ($2\times 10^{34}$
cm$^{-2}$sec$^{-1}$ for NLC/JLC, $3 \times 10^{34}$ cm$^{-2}$sec$^{-1}$ 
for TESLA) and lead to similar backgrounds from
beamstrahlung, pair production, and other machine-related
effects.  The physics case we will develop applies to both
schemes.  A decision between them must eventually be made on
the basis of cost, detailed technical advantages, and
upgradability, but we will not argue for either particular
approach in this report.

  The third issue concerns the ultimate upgrade of the energy of the $\ee$
collider to multi-TeV center of mass energies. Recent R\&D suggests that
this may be achievable.  It is likely that the needs of physics will
eventually call for experiments at such high energies, and so the collider
should be planned to support a program of successive energy upgrades. However,
the first stage of any program toward multi-TeV $\ee$ collisions will be   
a 500 GeV linear collider.  This first-stage machine now has a clear 
physics justification, and that will be the main focus of this report.  

\section{Lepton colliders and the long-term future of high energy physics}

The accelerators at CERN, Fermilab, DESY, and SLAC, which today
provide the highest energy particle collisions, were originally
envisioned and justified in an era when the fundamental
structures of the strong and weak interactions were completely
mysterious.  These facilities provided much of the data that
allowed these mysteries to be understood.  Through successive
upgrades and improvements, they also provided the data
that allowed the resulting theories to be tested with precision.
We have learned that with time, accelerators and individual
experiments outstrip predictions of their physics reach.  This
history implies that we should think about future accelerators
from a long-term perspective.  We begin this report with that
discussion.  Where may we expect to be, twenty years from now,
in our exploration of fundamental physics?  How can we get there?

\subsection{A twenty-year goal for high energy physics}

The beautiful experiments in particle physics over the
past twenty years have brought us to the point where we
are poised to discover the microphysical origin of mass.
In the Standard Model, the electroweak interactions are built
on the foundation of an $SU(2) \times U(1)$ gauge symmetry.
All of the mass terms in the Standard Model necessarily
violate this symmetry.  Masses can only appear because
some new fields cause this symmetry to be spontaneously
broken.

The spontaneous symmetry breaking cannot be explained
in terms of the known strong, weak, and electromagnetic
interactions.  In the 1980s, it was possible to believe
that the $W$ and $Z$ bosons were composite particles
\cite{bj,hungandsakurai,CFJ,suzuki}.  In the 1990s, when
electroweak radiative corrections were measured to be
in agreement with the $SU(2)\times U(1)$ gauge theory
\cite{Sirlin}, this possibility was  swept away.  At
the same time, the fundamental couplings of the strong,
weak, and electromagnetic interactions were precisely
measured.  At the weak interaction scale, these couplings
are too small to create a new state of spontaneously
broken symmetry.  Thus, the breaking of the electroweak
gauge symmetry must come from new fundamental interactions.
To explain the magnitude of the $W$ and $Z$ masses, these
interactions must operate at the TeV scale.

Over the next twenty years, a primary goal for high energy
physics will be to discover these new fundamental interactions,
to learn their qualitative character, and to describe them
quantitatively by new physical laws.  Today, although we can
guess, we do not know what form these laws will take.  It is
logically possible that the electroweak symmetry is broken
by a single Higgs boson.  More likely, the agent of symmetry
breaking will be accompanied by other new physics.  A popular
hypothesis is a supersymmetric generalization
of the Standard Model.  Other suggestions include models with
new gauge interactions, leading to a strongly-coupled theory
at TeV energies, and models with extra spatial dimensions and
quantum gravity at the TeV scale.

Aside from their own intrinsic importance, the study of these
new interactions will play a crucial role in our understanding
of the universe.  For example, supersymmetry is a theory of
space-time structure which requires modification of the theory
of gravity.  Other types of models, in particular those with
large extra space dimensions, necessarily invoke new space-time
physics at the TeV scale.

New physics is also needed to address one
of the mysteries of cosmology.   There is substantial evidence
that a large fraction of the total energy density of the universe
is composed of non-baryonic dark matter.  Recent estimates
require that dark matter should make up more than 80\% of  the total matter
in the universe \cite{darkmatter}.   A new stable particle with
a mass of about 100 GeV and an annihilation cross section of
electroweak size is an excellent candidate for this dark matter.
Models of electroweak symmetry breaking typically contain a particle
filling this description.  During recent years, an enormous amount
has been learned about the early universe, back to a time of about
1 second after the Big Bang, by the detailed comparison of primordial
element abundances with a kinetic theory of nucleosynthesis based
on measured nuclear physics cross sections \cite{KTbook}.   In
twenty years, we could have a precise knowledge of these new
interactions that would allow a predictive kinetic theory of the
dark matter.  This would push our detailed knowledge
of the early universe back to $10^{-12}$ seconds after the Big
Bang.

High energy physics has many concerns aside from the nature of
electroweak symmetry breaking.  The origin of the quark and lepton
flavors is mysterious; the pattern of masses and flavor mixings
is not understood.  The 
discovery that neutrinos have mass \cite{superK}
has added a
new dimension to this puzzle. In this decade, there will be a
significant effort, with contributions from many laboratories,
to measure the parameters of flavor mixing and CP violation.
These questions are all intimately related to the puzzle of
electroweak symmetry breaking.

There are two reasons for this.  First, in the Standard Model
all mass terms are forbidden by symmetry, and therefore all
masses, mixings, and CP violating terms must involve the
symmetry-breaking fields.  For  example, in a model in which
this breaking is due to fundamental Higgs bosons, the quark
and lepton masses, mixings, and CP violating angles originate
in the fermion couplings to the Higgs fields.  We will need
to know what Higgs bosons exist, or what replaces them, in
order to build a theory of flavor.  Second, deviations from
the conventional expectations for flavor physics are necessarily
due to new particles from outside the Standard Model.  If such
deviations are to be visible in the study of CP violation, for
example, the new particles must typically have masses of one to
several hundred GeV.  Given this mass scale, it is likely that
those particles are associated with the physics of electroweak
symmetry breaking.

Precision low energy experiments are designed to
search for deviations from the Standard Model.   Such deviations
indicate the presence of new particles which must be found at
high energies.  Models of new physics do not always predict such
deviations, and observed effects can be interpreted in multiple
ways.  So, there is no way to escape the need to search for new
particles directly at high energy.  In fact, we are {\em already} in
a situation where our current knowledge requires that new physics
be found at the next step in energy.  The need for new accelerators
can be seen from our study of the weak interactions, as a
consequence of the laws that we have established
experimentally in the past decade.

Thus, the elucidation of electroweak symmetry breaking should be the
key central goal for particle physics research in the next twenty
years.

\subsection{A twenty-year program for accelerators}

As we have just seen,
electroweak symmetry breaking requires new fundamental interactions;
it is our task to find and understand them.  In every example we
know of a fundamental law of Nature (with the possible exception
of Einstein's general relativity), the correct theoretical
understanding arose only with the accumulation of a large stock
of experimental data and the resolution of paradoxes within that
data.   New and varied experimental techniques were needed, both
to accumulate the basic data, and to crucially check or refute
intermediate hypotheses.

For the direct exploration of the TeV energy scale, only
two types of collision processes are feasible---proton-proton
and lepton-lepton reactions.  Proton-proton collisions have
the  advantage of very high center of mass energies and high rates.
However, this environment also has large backgrounds, mainly
from Standard Model gluon-gluon collisions.
Uncertainties from parton distributions
and from perturbative calculations limit the accuracy possible in
many precision measurements.  Lepton-lepton collisions have a
complementary set of advantages and disadvantages.  The
cross sections
are low, requiring high luminosity.  However,
new physics processes, if they occur, typically form a large fraction of
the total cross section.  Final states can be observed above
well understood backgrounds,
allowing unambiguous theoretical interpretation.
Cross sections for signal and background processes can be computed
to part-per-mil accuracy.  Lepton-lepton collisions provide
precise and model-independent measurements which complement those
from hadron machines.

It is well appreciated that, in developing our understanding of the strong
and electroweak interactions, proton and electron colliders made
distinct and complementary  contributions.  
As representative examples, recall
the discovery of nucleon and meson resonances, the $\Upsilon$,
and the $Z^0$ and $W^\pm$ at proton facilities and the corresponding
studies of deep inelastic scattering,  the charmonium and bottomonium
systems, the $Z^0$ resonance, and the $W^+W^-$ threshold at
electron machines.  In a natural evolution, results from $\ee$
have pointed to new processes in $D$ and $B$ meson decays which
have been probed further in high-rate hadron experiments.  In the
later sections of this report, we will discuss a number of specific
models that illustrate the way this complementarity might
play out at higher energies.

This logic leads us to plan, over the next twenty years, to study
the new interactions responsible for electroweak symmetry breaking
in both proton-proton and lepton-lepton collisions.  From our
experience with the strong and electroweak interactions, it is
likely that these new interactions will not be thoroughly understood
until we can look at them experimentally from energies above the
relevant particle masses.  In some supersymmetric models, it is
possible to stand above the whole spectrum at a center of mass
energy of 1 TeV.  But quite possibly---and necessarily for models
of strong-interaction electroweak symmetry  breaking---this
requires much higher energies, perhaps 5--10 TeV in parton-parton
collisions.

This challenge was the motivation for building the SSC.  With the
anticipated start of the  LHC experimental program in 2005, the
proton-proton program will at last begin. The LHC, operating at
14 TeV and a luminosity of $10^{34}$ cm$^{-2}$sec$^{-1}$,  has
parton collisions of sufficiently high energy that it is expected
to produce some signature of the new physics that underlies
electroweak symmetry breaking \cite{nolose, LHCnolose, SnowmassSC}.

For electron-positron colliders, all schemes for achieving high
energy collisions involve linear colliders.   The technology of
$\ee$ linear colliders is relatively new, but important expertise
was gained through operation of the SLC \cite{SLC}, which operated
at the $Z^0$ pole.   The natural next step for this technology is
a collider with 500 GeV center of mass energy.  A collider providing
this energy, and delivering the required luminosity, above $10^{34}$
cm$^{-2}$sec$^{-1}$, would be a critical step on the path toward
multi-TeV energies and very high luminosities.  At the same time,
as we shall see, a 500 GeV collider has sufficient energy to make
decisive contributions to the study of electroweak symmetry breaking.

The design of a 500 GeV linear collider must not preclude extension
to higher energies.  Indeed, both the  current 
warm  and superconducting
linear collider proposals explicitly include adiabatic extensions
to somewhat higher energies.  TESLA allows a stage of operation at 800 GeV.
The NLC/JLC plan includes ready expansion to 1 TeV and allows for
an upgrade to 1.5 TeV.  The pace of such an upgrade would depend
on the physics found at the LHC, as well as on results from the
first phase of 500 GeV operation.

In the context of a twenty-year plan, however, we must go even
further, and contemplate partonic collision energies of 5--10 TeV.
For hadron colliders, the VLHC program of R\&D now underway, or potential
upgrades to the LHC, could provide this; however it seems premature
to propose such a machine until the initial LHC results are  
available. A multi-TeV muon collider has received much recent attention, but 
there remain important R\&D issues to be resolved before its feasibility
can be determined.
In the past few years, a promising route to multi-TeV collisions has
emerged for $\ee$ colliders.
The possibility
of a 5 TeV $\ee$ linear collider was studied at Snowmass~'96
\cite{fiveTeV}, where three outstanding problems were
identified:  the lack of a feasible RF power source for high
frequency accelerating structures, the large length of the final
focus sections, and the tight manufacturing and alignment tolerances
for the accelerating structures.  Since then, there has been
considerable progress.   A major rethinking of the two-beam
(CLIC) acceleration scheme makes this concept, in which a low-energy,
high-current beam is used to generate high-frequency RF, look
promising as a power source for very high energy acceleration
\cite{CLIC}.  Indeed, such schemes now look feasible for lower
RF frequencies (for example, at X band), and this could
provide a natural evolution path to higher accelerating gradients
\cite{RRuth}.  New compact final focus layouts \cite{Pantaleo}
have been recently incorporated into the NLC design.

The issue of manufacturing and alignment tolerances is central
to the successful operation of any high-luminosity linear
collider.  This issue is presented in a more manageable form
in the design of a 500 GeV collider with either warm or
superconducting  RF.  Moreover, the experience of building
and running this machine will be
an invaluable prerequisite to eventual $\ee$ experimentation at
multi-TeV energies.  In addition, any multi-TeV $\ee$ linear
collider will be placed in a long, straight tunnel exactly like
the one on the site of a 500 GeV machine and perhaps could reuse
the damping rings and injector complex of the 500 GeV stage.
Thus, a 500 GeV linear collider is
the first stage of a twenty-year exploration in $\ee$ physics.

\section{Parameters of a 500 GeV linear collider}

The designs of linear colliders have evolved dramatically over
the past five years, based in part on experience from the SLAC
Linear Collider operating at 91 GeV, and in part on extensive
collaborative R\&D efforts in Europe, Japan and the United
States.  At this
writing, the machine parameters are still being evaluated; this
section is intended to give the currently envisioned scope of
the possible accelerator projects.

The TESLA collider, developed by a collaboration led
by DESY, would employ superconducting RF
accelerating cavities operating in L-band (1.3 GHz).  The
JLC (KEK) and NLC (SLAC, LBNL, LLNL, FNAL) designs are
based on warm accelerating
structures operating in X-band (11.4 GHz).  Initial construction
of each of these is expected for a 500 GeV machine.  A variety of
important differences in the designs follow from the basic choice
of accelerating frequency. (KEK is  also considering a
C-band variant operating at 5.7 GHz.)

\begin{table}
\begin{center}
\begin{tabular}{|c||c|c|} \hline
 ~                            &  TESLA          & NLC/JLC \\ \hline
$E_{{\rm CM}}$ (GeV)                &  500            & 500     \\
RF frequency (GHz)            &  1.3             & 11.4    \\
Repetition rate (Hz)                 &  5             & 120     \\
Luminosity ($10^{34}$ cm$^{-2}$sec$^{-1}$)   &  3.4
&
2.2
\\ Bunch separation (ns)         & 337         & 1.4     \\
Effective gradient (MV/m)     &  22             & 50.2    \\
Beamstrahlung (\%)            &  3.3           & 4.6     \\
Linac length (km)            &  31             & 10.8    \\ \hline
\end{tabular}
\caption{\small Basic parameters of the high-luminosity TESLA and
NLC/JLC accelerator designs.}
\end{center}
\label{tab:machines}
\end{table}

The main parameters of TESLA and the X-band NLC/JLC
are shown in Table~1.
For all proposals, electron beam polarization of 80\% is
expected.  Production of polarized positrons can be envisioned
by creating polarized photons in sophisticated undulator magnets,
or by backscattering polarized high-power lasers, but these
possibilities require further development.  In all proposals,
the collider can also be operated for $e^-e^-$ collisions with
some loss in luminosity.  By backscattering laser beams, it
may be possible to create a high-luminosity gamma-gamma collider
with a center of mass energy of about 80\% of that for $\ee$.

The U.S. design of the NLC underwent a DOE readiness review to
initiate the Conceptual Design Report in May 1999.  The Review
Committee was positive in its assessment of the technical
design.  The cost was estimated at \$7.9B.  After subtraction
of contingency, escalation, and
detectors, these costs were distributed over the major subsystems
as follows: injectors (19\%), main linacs (39\%), beam delivery (11\%),
global costs (17\%), management/business (14\%).  The DOE
decided not to proceed with the official CD-1 milestone in view
of this cost.  Present work is focused on cost and possible
scope reductions.    In the past year,
progress has been made in identifying areas of
savings, including the use of permanent magnets for the beam lines,
electronics distributed along the linacs, modifications to the
injectors, and considerable reduction of the length of the final focus.
Demonstrated improvements in the klystrons and modulators should give
a reduction of RF power costs.   Taken together, these developments
are estimated to reduce the cost by 30\%.  Scope reductions, including
building the linacs initially for 500 GeV operation, with subsequent
civil construction for higher energy, could yield a further 10--15\%
reduction in the initial cost.

The luminosity expected for the NLC design depends critically on the
precision with which one can build and align the disk-loaded 
accelerating structures of the main X-band linac.
Recent tests have demonstrated
that structures can be produced with 2--3 times better accuracy than
projected in the 1999 review, and that monitors built 
into these structures can 
measure their position with respect to the beam to within a few microns.
Re-examination of the beam parameters in the light of these results has led
to the realization that the luminosity of the collider can be expected to 
be 3--4 times higher than projected in 1999, although it is likely to 
require  some
period of running to carry out the needed beam-based alignment of 
the accelerator.  It is reasonable to assume that the collider will begin 
operation at $5\times 10^{33}$\lunit \  and that, over a period of time, 
it will reach
the design luminosity of $2.2\times 10^{34}$\lunit\ shown in Table 1.  This
would yield 100 fb$^{-1}$ of accumulated data in the first year of operation
and 200 fb$^{-1}$/yr in subsequent years.

Each of these proposals includes possible adiabatic upgrades
in energy.  The TESLA collider can be expanded to 800 GeV through
higher accelerating gradients.   The NLC/JLC energy upgrade to
1 TeV could be achieved through an increase in the linac lengths
and the addition of more RF structures.  Improvements in RF gradients
or further increases in length
could allow operation at 1.5 TeV.  It is important for the long
term evolution of the linear collider that the flexibility to
implement these options be included in the initial machine design.

Work has been done at CERN (CLIC) to develop the RF power for
acceleration to even higher energies.  The idea is to generate
wakefield power for the main linacs using a high current, low
energy drive beam operating  at low (L-band) frequencies.
Recent work at SLAC has expanded this concept to
incorporate a recycling drive beam train that is cheaper, more
compact and efficient than the original CLIC concept.
Accelerating gradients of about  100 MV/m
are envisioned for this two beam design.   The two beam linear
collider offers an attractive possibility for later expansion
of the linear collider to multi-TeV operation, and suggests the
potential for an evolving accelerator facility that can follow
the initial phase of physics results.  Recent R\&D suggests that
the use of the two beam drive technology is as well suited for
linacs operating in the X-band as for the 30 GHz structures
originally envisioned by CLIC, although the limits to feasible
gradients are not clear.

For the NLC design with permanent magnets in the beam
lines, the energy for operation cannot be decreased
below half its maximum.   As discussed in the next
sections, physics considerations may dictate that a wider
range of energies is needed.  In particular, a return to the
$Z^0$ pole may be desirable to improve the precision of the
electroweak measurements.   Similarly, if the Higgs boson
is in the low mass region favored by the Standard Model or
supersymmetry, it may be advantageous to accumulate substantial
integrated luminosity at the energy of the maximum Higgs cross
section and, at the same time, explore the high energy region.
Recently, consideration has been given to providing a second
beam operating at lower energies.   This beam would be extracted
from the main accelerator and accelerated in unused time slices
of the AC duty cycle.  The extra power needed for this operation
could be low because of  the reduced energy of the beams.  Low and
high energy beams would be delivered to dedicated detectors
installed at separate interaction points in the beam delivery
region.

\section{Why we expect new physics below 500 GeV}

At Snowmass '96, it was argued that a 1.5 TeV $\ee$ collider
is roughly equivalent to the LHC in its ability to
detect the new physics  related to electroweak symmetry
breaking \cite{SnowmassSC}.  However, this point will certainly
be moot by the time such a linear collider operates.  The
real question that we must address is different:
{\em In an era in which the LHC is already exploring the new
interactions responsible for electroweak symmetry breaking,
what critical information must $\ee$ experiments add, and at
what $\ee$ center of mass energies should this information be
sought?}

Today, there is considerable evidence that an $\ee$ collider
program should begin at a center of mass energy of 500 GeV.
This evidence is indirect and will remain so until the new
particles responsible for electroweak symmetry breaking are
discovered.  The case rests on the large body of precision
data acquired over the past ten years.
These data agree remarkably with the minimal Standard Model.  When
interpreted using this model, they require that the Higgs boson be light.
The data also place strong constraints on possible new physics associated
with electroweak symmetry breaking.  These constraints define distinct 
pathways for new physics which will be tested at the next generation of 
colliders.

Following the guidance of the precision data, we will argue in this  
section
that a 500~GeV linear collider will be needed whatever the outcome of the 
LHC experiments might be.  In Sections 4.1--4.3, we will outline why
there should be a light Higgs boson with mass below about
200 GeV.  In Section 4.4, we will argue that, if the new physics
includes supersymmetry, the lightest superpartners should be
found at a 500~GeV collider.  There are known ways to evade
these arguments, but they too give rise to crucial tests in
$\ee$ collisions at 500~GeV, as we will discuss in Section
4.5.  Finally, in Section 4.6, we will address the question:
what if the LHC sees no new physics?

\subsection{A fundamental versus composite Higgs boson}

Models of electroweak symmetry breaking divide into two groups
at the first step.  Is the symmetry breaking induced by a
fundamental scalar field or by a composite object?  Is electroweak
symmetry breaking a weak-coupling phenomenon, or does it
require new strong interactions?  These basic questions
have driven the study of electroweak symmetry breaking for twenty
years \cite{Weinberg,Susskind}.   Many
people use analogies from QCD or superconductivity to argue against the
plausibility of fundamental scalars, or use the perceived beauty
of supersymmetry to motivate a fundamental scalar Higgs field.
We believe that it is possible to make a preliminary judgment---in
favor of a fundamental Higgs field---on the basis of the data.
This will be important, because models in which the Higgs is
fundamental favor a light Higgs boson, while other models favor
a heavy Higgs resonance, or none at all.

The simplest model of electroweak symmetry breaking is the minimal  
version
of the Standard Model, which introduces one elementary
Higgs field and nothing else.
This model is consistent with the present data, but it is totally  
inadequate
as a physical theory.  In this model, the mass parameter $m^2$
of the Higgs field is a free
parameter which cannot be computed  as a matter of principle, because it 
receives an infinite additive renormalization.  Electroweak symmetry is 
broken or not according to whether this parameter, after renormalization,
is positive or negative.
  If the infinite radiative corrections are made finite by a
cutoff at some energy $M$, $m^2$
can be much less than $M^2$ only if the radiative corrections are finely
tuned to cancel.  If $M$ is taken to be the Planck  scale,  these
corrections must cancel in the first 30 decimal places.  Theorists often
consider this to be a problem in its own right (the `gauge hierarchy 
problem').  This problem is a symptom of the fact that the
Standard Model is only a parametrization, and not an explanation,
 of electroweak symmetry breaking.

Theories of electroweak symmetry breaking can be constructed either with 
or without fundamental Higgs particles.  The preference we have
expressed for a fundamental Higgs particle is reflected in the history
of the subject.
Phenomenological models of supersymmetry introduced in the early
1980s \cite{Inoue,Ibanez,Ellis,AlvarezG} are as valid today as
when they were first created. 
On the other hand, the predictions of the early dynamical models
 (as reviewed, for example, in \cite{LanePeskin}) have been
 found to be inconsistent with experiment, requiring major changes
 in model-building strategies.

To discuss this point, we must define what we mean by a `fundamental
scalar field'.  A particle which looks fundamental and structureless
on one length scale can be seen to be composite on a smaller length
scale.  In nuclear physics, and more generally in scattering processes
with energies of a few hundred MeV, the pion can be treated as a
structureless particle. However, in hard QCD processes, the pion must
be treated as a quark-antiquark bound state. At the other extreme,
string theory predicts that even quarks and leptons have a finite
size and an internal structure at the Planck scale.  In almost any
theory, a particle can at best be considered fundamental at some
particular distance scale.  The question here is whether the Higgs
boson is elementary well above the scale of the new interactions
responsible for electroweak symmetry breaking.  In the following
discussion, we use the term `fundamental Higgs' for the case that
there is a scalar Higgs field in the Lagrangian at an energy scale
of 20 TeV.

The answer to this question has direct implications for the theory
of the  quark and lepton masses.  These masses arise through $SU(2)
\times U(1)$ symmetry breaking, from terms in the effective Lagrangian
that couple left-handed to right-handed fermions.  If there is a
fundamental Higgs field, a typical term has the form
\beq
          \delta \L =    \lambda_f  \bar f_L \phi f_R  +  {\rm h.c.} \ ,
\eeq{Higgstomass}
where $\phi$ is an $SU(2)$-doublet Higgs field and the coupling
$\lambda_f$ is dimensionless.  The fermion $f$ obtains mass when
$\phi$ acquires a vacuum expectation value.    To explain the size
of the mass,  a theory must contain new interactions that fix
the value of $\lambda_f$.  Because $\lambda_f$ is dimensionless,
these interactions can occur, without prejudice, at any energy
scale larger than 20 TeV.  In typical models with a fundamental
Higgs boson, these interactions occur at the scale of grand
unification, or even above.

If there is no fundamental $SU(2)$-doublet scalar field, the
interaction \leqn{Higgstomass} does not exist.  Instead, one
must write a more complicated interaction that couples $\bar
f_L f_R$ to other new fields.  For example, in technicolor
models, one writes
\beq
   \delta \L =   {g^2\over M_E^2} \bar f_L f_R  \bar Q_R Q_L  + {\rm  
h.c.} \ ,
\eeq{Qtomass}
where $Q$ is a new heavy fermion with strong interactions at
the TeV scale.  This is a dimension-6 operator, and therefore
we have written a coefficient with the dimensions (mass)$^{-2}$.
If the operator $(\bar Q_R Q_L)$ acquires a vacuum expectation
value at the TeV scale and this operator is expected to generate
a 1 GeV fermion mass, $M_E$ must be roughly  30 TeV.  The
four-fermion operator \leqn{Qtomass} can be induced by the
exchange of a heavy boson of mass $M_E$.  However, whatever the
mechanism that leads to this operator, the physical interactions
responsible must operate at some energy scale not too far above
$M_E$.  This means that, unlike the previous case, the interactions
that determine the quark and lepton masses and mixings must occur
at energies not so far above those we now probe experimentally.

In fact, these interactions must occur at sufficiently low energies
that they would be expected to contribute significantly to $\mu \to
e\gamma$ and $K\to \mu e$, and to $K$--$\bar{K}$, $B$--$\bar B$,
and $D$--$\bar D$ mixing.  The fact that these processes are not observed is a
severe problem for dynamical theories.  A further problem arises
from the large size of the top quark mass. To  produce a mass as
large as is observed, the mass scale $M_E$ for the top quark---and,
by symmetry, for the $b_L$---must be close to 1 TeV.   This new
interaction would be expected to lead to enhanced flavor-changing
neutral current amplitudes, and to few-percent corrections to the
$Zb\bar b$ coupling \cite{CSS}.

These experimental observations have eliminated essentially all
simple models of dynamical symmetry breaking.  The only models
that survive have complex new dynamics (\eg, \cite{randallGIM,
Sundrum,Apostmodern}) or, below energies of several TeV, behave
almost exactly like the Standard Model with a scalar Higgs field
(\eg, \cite{TCseesaw}).  Neither type of model resembles the
attractive intuitive picture that first led people to explore
electroweak  symmetry breaking by new strong interactions.

Generalizations of the simplest Standard Model with additional
fundamental scalar fields have also been proposed.  But these
have little motivation, and like the minimal Standard Model,
the Higgs vacuum expectation value, and even the existence of
electroweak symmetry breaking, cannot be predicted as a matter
of principle.  

The simplest models with a fundamental Higgs
field in which electroweak symmetry breaking results from a
calculation, rather than a parameter choice, are those with
supersymmetry.  Without debating the virtues or deficits
of supersymmetric models, what is relevant here is that
supersymmetric models have not been significantly constrained
by the  precise experimental measurements of the past twenty
years.  Supersymmetric particles give very small effects in electroweak
precision measurements because the masses of the superparticles
preserve $SU(2) \times U(1)$ gauge symmetry, and so do not
require electroweak symmetry breaking.
In models that decouple
in this way, new particles with mass $M$ give corrections to
the Standard Model predictions at the $Z^0$
which are of size
\beq
             {\alpha\over \pi} { m_Z^2 \over M^2}  \ .
\eeq{Zcorrect}
As long as we stay below the energy at which the new particles
actually appear, their influence is very small.  Then, as we
pass the threshold, new physics appears suddenly.   Supersymmetry
thus naturally suppresses deviations from the Standard
Model---until we begin to produce the supersymmetric particles.
Models with dynamical electroweak symmetry breaking almost always
contain 
heavy matter states which have chiral couplings and thus do not decouple
from electroweak symmetry breaking.
In these models, one expects significant
corrections to the Standard Model well below the energy scale
of the new particles.

In addition to this decoupling, the early supersymmetry models
made two important predictions. The first was that the top
quark mass should be heavy.  This tendency arises from the
fact that, in supersymmetric models, electroweak symmetry
breaking can be  triggered by radiative corrections due
to the top quark Yukawa coupling.  The papers \cite{Inoue,
Ibanez,Ellis,AlvarezG} all quoted lower bounds on the top quark
mass, ranging from 50 to 65 GeV.  (Later, corners of parameter
space were found in which the top quark mass could be lower.)
Supersymmetry
readily accomodates a top quark mass as large as 175 GeV.
The second prediction was that the value of $\sstw$ should
be close to 0.23 (as now observed), rather than the value 0.21
preferred in the early 1980's.  This prediction arises from
grand unification with the renormalization group equations of
supersymmetry \cite{DWR,EinandJ,IbanR}.  The precise
determination of $\alpha_s$ and the electroweak couplings
at the $Z^0$ has given even stronger support to the idea of
supersymmetric grand unification, with the issue now at the
level of detailed higher-order corrections \cite{LPol}.

Of course it is premature to make a final decision between
the different models.  For this, we must discover and study
the Higgs boson, or whatever takes its place.  But, in planning
where we should look for these phenomena, we should take into
account that models with fundamental Higgs bosons passed the
first tests presented by the data, while the early dynamical
models did not.

\subsection{A fundamental Higgs boson should be light}

In the previous section, we noted that in models with
fundamental Higgs bosons, the Higgs is typically light.  In
this section, we will quantify that statement with upper
bounds on the Higgs mass.

In the Standard Model, the mass of the Higgs boson is determined
in terms of the Higgs field expectation value $v$ and the Higgs
self-coupling $\lambda$ by the relation
\beq
            m_h = \sqrt{2\lambda} v \ ,
\eeq{mhval}
with $v = 246$ GeV determined by  the values of the $W$ and $Z$
masses.  A bound on $\lambda$ thus implies a bound on $m_h$.  For
example, $\lambda < 1$ implies $m_h <  350$ GeV.  How large
can $\lambda$ reasonably be?

Like $\alpha_s$, $\lambda$ is a running coupling constant, but
in this case radiative corrections drive $\lambda$ to larger values
at higher energies.  Just as the running $\alpha_s$ diverges at
$\Lambda_{\msb}$, signaling the onset of nonperturbative QCD
effects, the running $\lambda$ diverges at a  high energy scale
$\Lambda_h$.  Presumably, this must signal the breakdown of the
fundamental Higgs picture.   The relation between $\Lambda_h$
and  the value of $\lambda$ at the weak interaction scale can
be computed from the Standard Model \cite{basicmh}. It is conveniently
written, using \leqn{mhval}, as
\beq
              m_h =  {1000 \ \mbox{GeV}\over \sqrt{\ln (\Lambda_h/v)} }
\eeq{mhlimit}
The value of $m_h$ in \leqn{mhlimit} is the largest Higgs boson
mass compatible with a Higgs field which is elementary at the
scale $\Lambda_h$.  For  $\Lambda_h = 20$ TeV, $m_h < 500$ GeV.

A much stronger limit on $m_h$ is obtained if one takes seriously
the  experimental evidence for grand unification and assumes that
the Higgs boson is a fundamental particle at the grand unification
(GUT) scale.  If we naively put  $\Lambda_h > 10^{16}$ GeV into
\leqn{mhlimit}, we find $m_h < 180$~GeV.  Successful
grand unification requires supersymmetry and brings in ingredients
that make the computation of $m_h$ more complex.  But, detailed
analysis of supersymmetric grand unified models has shown that the
idea of an upper bound on $m_h$ remains valid. In 1992, two
groups presented systematic scans of the parameter space of
supersymmetric grand unified theories, demonstrating the bound
$m_h < 150$ GeV \cite{OM,Kane}.  Exceptions to this constraint
were later found, but still all known models satisfy $m_h <
205$ GeV \cite{QE}.

The Minimal Supersymmetric Standard Model is a special case.  In
this model, the tree-level potential for the lightest Higgs boson
is determined completely by supersymmetry.  Radiative corrections
to this potential are important. Nevertheless, it can be shown that
$m_h < 130$ GeV in this model \cite{MSSMhiggs}.  Here the conclusion
is independent of any assumptions about grand unification.

\subsection{The constraint on the Higgs mass from precision  
electroweak data}

The previous two sections did not make any
reference to the determination of the Higgs boson mass from the
precision electroweak data. Those data give a second, independent
argument for a light Higgs boson.  The Higgs field contributes to
electroweak observables through loop corrections to the $W$ and $Z$
propagators. The effect is small, of order $\alpha \ln (m_h/m_W)$,
but the accuracy of the measurements makes this effect visible.  A
fit of the current data to the Standard Model, using the measured
value of the top quark mass, is consistent only if $\ln(m_h/m_W)$
is sufficiently small.  The LEP Electroweak Working Group finds
upper limits $m_h < 188$ GeV at the 95\% CL and $m_h < 291$
GeV at the 99\% CL \cite{newLEP}.  
  Even using more conservative estimates of the
theoretical errors \cite{OkunZloop}, the limit on the Higgs boson mass is
       well within the range of a 500 GeV $\ee$ collider.

This Standard Model limit does not obviously apply to more general
models of electroweak symmetry breaking.  In what follows we will
discuss its validity in various models.  As previously, the
result depends on whether or not the Higgs is fundamental.

We have noted in Section 4.1 that models with a fundamental Higgs
boson typically satisfy decoupling.  The practical effect of this
is that, if new particles are sufficiently massive that they cannot
be produced at LEP 2, their contributions to electroweak corrections
are too small to affect the current global fits. In particular, fits 
to models of supersymmetry produce upper bounds on the
Higgs mass similar to those from the Standard Model.

It is difficult to make a model with dynamical electroweak symmetry
breaking that is consistent with precision electroweak measurements.
The simplest technicolor
models, for example, give several-percent corrections to electroweak
observables \cite{holdomT,randall,pandt}; effects this large are
completely excluded.  Even models with one $SU(2)$ doublet of
techni-fermions give corrections of a size roughly double that
for a 1000 GeV Higgs boson.  With  models of this type, it is
typically necessary to invoke some mechanism that compensates the
large corrections that appear in these models, and then to adjust
the compensation so that the precision electroweak constraint
is obeyed.  In this process, the constraint on the Higgs boson
mass can be relaxed.

A recent review \cite{WP} describes the three different compensation
strategies that have been presented in the literature.  One of
these strategies leads to a lower value of the $W$ mass and a
larger $Z$ width than predicted in the Standard Model.  It can be
distinguished by the improved precision electroweak measurements
that we describe in Section  5.6.  The other two strategies predict
either new light particles with electroweak charge or other
perturbations of Standard Model cross sections visible below 500 GeV.
Thus, models based on new strong interactions can avoid having
Higgs bosons below 500 GeV, but they predict phenomena observable
at a 500 GeV linear collider.

\subsection{The lightest supersymmetry partners are likely to appear  
at 500 GeV}

For supersymmetric models of electroweak symmetry breaking, the
arguments of the previous two sections give us confidence that we
will be able to produce the lightest Higgs  boson.  But we also
need to study the supersymmetry partners of quarks, leptons, and
gauge bosons.  Thus, we must also explore how heavy these particles
are likely to be.

Because supersymmetric generalizations of the Standard Model revert
to the Standard Model when the superpartner masses are taken to be
heavy, it is not possible to obtain upper limits on the masses of
supersymmetric particles by precision measurements.  One must take
a different approach, related to the problems of the
Standard Model discussed at the beginning of Section 4.1.  As we
noted there, it is a property of the Standard
Model that  radiative corrections from a high mass scale $M$ contribute 
additively to the Higgs mass and vacuum expectation value,
affecting $m_W$ in the form
\beq
          m_W^2 =     {g^2 v^2\over 4} + {\alpha\over \pi} M^
2 + \cdots \ .
\eeq{additiveM}
It is possible to obtain a value of the $W$ mass much less than $M$
only if the various contributions cancel to high accuracy.  For example,
these terms must cancel to 3 decimal places for $M = 20$ TeV or to 30
decimal places for $M = 10^{18}$ GeV.  Supersymmetry solves this problem
by forbidding such additive corrections to $m_W^2$.  But this restriction
applies only if supersymmetry is unbroken. If the masses of the
superpartners are much greater than $m_W$, the fine-tuning problem
returns.

\begin{table}
\begin{center}
\begin{tabular}{|l|c|c|c|c|}\hline
 & $\ch 1$ & $\s g$ & $\s e_R$ & $\s u$, $\s d$  \\[.5ex]
\hline\hline
Barbieri-Giudice \cite{BG}    &    110   &   350  &   250  &
420  \\
Ross-Roberts \cite{Ross}      &    110   &  560   &   200  &
520   \\
de Carlos-Casas  \cite{deCarlos}&  250   &  1100  &  450   &
900  \\
Anderson-Castano \cite{ACast} &     270  &  750   &  400   &
900   \\
Chan-Chattopadhyay-Nath \cite{ChanNath}&  250   &   930  &  550
&  900 \\
Giusti-Romanino-Strumia \cite{Giusti}  &  500   &  1700   &  600
& 1700   \\
Feng-Matchev-Moroi \cite{naturalFeng}
         &  240/340   &  860/1200 &  1700/2200  &  2000/2300
\\[.5ex]
\hline
\end{tabular}
\caption{\small Upper limits on supersymmetry particle masses (in GeV)
from the fine-tuning criterion found by various groups.
In the last line, we have chosen two different breakpoints
in fine-tuning from the results given in the paper.}
\end{center}
\label{tab:natural}
\end{table}

This theoretical motivation leads us to expect that supersymmetric
particles are most natural if they are light, of order a few hundred
GeV.  One can try to quantify this argument by limiting the amount
of accidental cancellation permitted in the calculation of $m_W$.
By now, many authors have studied this cancellation in a variety
of supersymmetric models.  In Table~2, we show
the upper limits on supersymmetry
particle masses found by seven groups for the parameter space of
minimal gravity-mediated supersymmetry models (m\-SUGRA). The
detailed calculations leading to these limits are different and,
in many cases, involve conflicting assumptions.  These differences
are reflected in the wide variation of the limits on first- and
second-generation slepton and squark masses evident in the table.

Nevertheless, these analyses are in general
agreement about the required scale of the gaugino masses and (except for
\cite{Giusti}) expect chargino pair production to be kinematically
accessible at or near 500 GeV.   A simplified but quantitative
argument for this bound can be made \cite{naturalFeng} by writing
the expression for $m_W^2$ in terms of  the underlying parameters of
the model, and eliminating these in terms of physical particle masses.
For the representative value  $\tan\beta = 10$, one finds
\beq
   m_W^2 = - 1.3 \mu^2 + 0.3 m^2(\s g)  +   \cdots  \ ,
\eeq{Wcancellation}
where the terms displayed involve the supersymmetric Higgs mass
parameter and  the gluino mass.  The omitted terms involving scalar
masses are more model-dependent.  The gluino mass enters through its
effect on the renormalization of the stop mass.  For a gluino mass
of 1 TeV, the requirement that the $W$ mass is no larger than
80 GeV requires a fine-tuning of 1 part in 50.  A similiar level
of fine-tuning is needed if $\mu$ is greater than 500 GeV.

As we will discuss in Section 5.2, the masses of the two charginos
are closely related to the wino mass parameter $m_2$ and the Higgs
mass parameter $\mu$.  In particular, the lighter chargino mass lies
close to the smaller of these two values.  The parameter $m_2$
is connected to the gluino mass in mSUGRA models by the grand
unification relation
\beq
m_2/ m(\s g) \approx \alpha_w/\alpha_s \approx 1/3.5  \  .
\eeq{choverg}
This relation also holds in gauge-mediation, where, in addition,
the masses of sleptons are predicted to be roughly the same size
as the mass of the chargino.  In other schemes of supersymmetry
breaking, the chargino/gluino mass ratio can differ; for example,
in anomaly-mediation, $m_2/ m(\s g) \approx 1/8$.   In all of
these models, the bound on $m(\s g)$ implies a strong bound on
the lightest chargino mass.  The fact that  both $m_2$ and $\mu$
are bounded by the fine-tuning argument implies that there is
also a bound on the mass of the heavier chargino.  Indeed, one
typically finds that the full set of chargino and neutralino
states can be produced at an 800 GeV $\ee$ collider
\cite{naturalFeng}.

Although the fine-tuning limits are by no means rigorous, they indicate
a preference for light supersymmetry partners. 
They encourage us to expect that we will be able to study the
 lighter chargino and neutralinos at the initial stage of the
linear collider program, and all gauginos with a modest upgrade of the      
 energy.

\subsection{What if there is no fundamental Higgs boson?}

Despite our arguments given in Section 4.1
for preferring a fundamental Higgs boson,
electroweak symmetry breaking could result
from a new strong interaction.
Whereas for supersymmetry we have a well-defined minimal model, albeit
one with many
free parameters, here even the basic structure of the model is
unknown and we will need more guidance from experiment.  It is
thus important to identify measurements that probe possible new
strong interactions in a variety of ways.

In models with a composite Higgs boson, the Higgs mass can be large,
500 GeV or higher.  If the Higgs is very heavy, there is no distinct
Higgs resonance.  A heavy but narrow Higgs boson can be studied at the
LHC in its $Z^0Z^0$ decay mode, and at a higher energy $\ee$ collider.
A broad resonance or more general new strong interactions can be
studied through $WW$ scattering at TeV energies.  This study can also
be done at the LHC and at a higher energy linear collider \cite{SnowmassSC}.
However, in this case, the experiments are expected to be
very challenging.  Certain classes of models which are preferred by
the arguments of Section 4.1 (\eg, \cite{TCseesaw}~)
 predict that no effect will
be seen in these reactions.

 In view of this, it is essential to have another way to probe models
with a composite Higgs boson.  This can be done by studying the effects of the
new physics on the Standard Model particles that couple most strongly
to it---the $W$, $Z$, and top quark. 
Because the $Z$ couples to light fermions through a gauge current, effects
of the new strong interactions are not expected to appear in $Z$ decays,
except possibly in $Z\to b\bar b$.  The first real opportunity to observe
these effects will come in the study of the $W$, $Z$, and $t$ couplings.
Effects of strong-interaction electroweak symmetry breaking can  
appreciably
modify the Standard Model predictions for these couplings.

Without a specific model, it is difficult to predict how large these  
effects
should be, but some estimates provide guidance.  For example,
triple gauge boson couplings can be related to parameters of the effective
chiral Lagrangian describing the nonperturbative $SU(2)\times U(1)$  
symmetry
breaking.  The parameter
$\Delta \kappa_\gamma$ which contributes to the $W$ anomalous  
magnetic dipole
moment, is given by \cite{SnowmassSC}
\beq
      \Delta \kappa_\gamma =  - 2 \pi \alpha_w (L_{9L} + L_{9R} +  
L_{10}) \ ,
\eeq{forkappa}
where the $L_i$ are dimensionless parameters analogous to the  
Gasser-Leutwyler
parameters of low energy QCD \cite{GasserL}.  Naively putting in the
QCD values, we find
\beq
     \Delta  \kappa_\gamma   \sim   - 3 \times 10^{-3} \ .
\eeq{kappaest}
A deviation of this size cannot be seen
at LEP or the Tevatron. It is close to the
expected error from the LHC.  However, a 500 GeV $\ee$ collider can
reach this sensitivity by the precision study of $\ee\to W^+W^-$, as
we will discuss in Section 5.5.

For the top quark, somewhat larger effects are expected, specifically
in the $Z \bar t t$ coupling.  As we noted in Section 4.1, it is
already a problem for these models that the decay width for $Z\to b
\bar b$ agrees with the Standard Model.  However, models can contain
several competing effects which add destructively in the $Z \bar b b$
coupling but constructively in the  $Z\bar t t$
coupling \cite{CST,HagiTT,Mahanta}.
In that case, 5--10\% corrections to the $Z \bar t
t$ coupling would be expected.  These would produce corrections to the
cross section for $\ee\to t\bar t$ which would be observed through the
measurement of this cross section at a 500 GeV $\ee$ collider.  We
will discuss the program of precision measurements of anomalous top quark
couplings in Section 5.3.

In the past few years, there has been a theoretical preference
for supersymmetry and other weakly-coupled models of electroweak
symmetry breaking.  If supersymmetric particles are not discovered
at the LHC, this situation will change dramatically.  In that case,
anomalous $W$ and $t$ coupling measurements at an $\ee$ collider
will be among the most central issues in high-energy physics.

\subsection{What if the LHC sees no new physics?}

Though we expect that the LHC will reveal a rich spectrum of new
particles, it is possible that the LHC will see no new phenomena.
How could the LHC see no sign of the interactions responsible for
electroweak symmetry breaking?  The LHC should not fail to find
supersymmetry if it exists.  The LHC, at full luminosity, should
be sensitive to resonances in $WW$ scattering beyond the limit
set by $s$-channel unitarity.  Thus, if the  LHC fails to find
signatures of electroweak symmetry breaking, it will not be
because this collider does not have high enough energy.  The
scenarios in which the LHC fails---which, we  emphasize, are very
special scenarios occupying a tiny volume of  typical parameter
spaces---are those in which there is a light Higgs boson that
does not have the decay modes important for detection at the LHC.

A Higgs boson with mass larger than about 150 GeV has a large
production cross section from $WW$ fusion and a substantial
branching ratio to decay back to $WW$.  Even if the $hWW$
coupling is diluted as described below, it is hard for us to
imagine that this signature will not be seen at the LHC.

But for Higgs bosons with mass below 150 GeV, it is possible
that there are new particles with masses tuned so that their
loop contributions to the $h\gamma\gamma$ coupling cancel the
Standard Model contribution.  This can happen, for example,
at specific points in the parameter space of the Minimal
Supersymmetric Standard Model \cite{WMKane}.  It is also possible
that a substantial fraction of the Higgs decays are to invisible
final states such as $\neu1 \neu1$.  Finally, if there are several
neutral Higgs fields, each of which has a vacuum expectation
value, the strength of the squared $hWW$ coupling for any individual
field will be divided by the number of fields participating.  Any
of these three possibilities would compromise the ability of the
LHC experiments to find and study the Higgs boson.   The ability
of an $\ee$ collider to see the Higgs boson does not depend on
the Higgs decay pattern, but only on measurement of missing mass
recoiling against a produced  $Z^0$ boson.   Thus, a 500 GeV
$\ee$ collider would be the ideal instrument to study the Higgs boson
under these special circumstances, as discussed in Section 5.1.

There is another way that the LHC could `discover nothing' which we
must confront.  It could be that the Standard Model is correct
up to a mass scale above $10^{16}$ GeV, and that the only new
physics below that scale is one standard Higgs boson.  This conclusion
would be extremely vexing, because
it would imply that the reason for the spontaneous breaking of
electroweak symmetry and the values of the quark and lepton masses
could not be understood as a matter of principle.  In that case, before
giving up the quest for a fundamental theory, we should search in
detail for non-standard properties of the observed Higgs boson.
We will show in Section 5.1 that this study is ideally done at an
$\ee$ linear collider.  In this scenario, the mass of the Higgs boson
must lie in a narrow window between 140 and 180 GeV, so an energy of
500 GeV  would be
sufficient.  The final confirmation of the Standard Model would be  
compelling
only after the Higgs boson has passed all of the precision tests
possible at an $\ee$ collider.

\section{Physics at a 500 GeV linear collider}

We have argued in the previous section that there is a high
probability that new physics associated with electroweak symmetry breaking
will appear at a 500 GeV $\ee$ collider.  We have given two different 
arguments that the Higgs boson should appear in $\ee$ annihilation at this
energy. For models with TeV-scale supersymmetry, it is likely that the 
lighter chargino and neutralino states can also be found.  For models with
strong-coupling electroweak symmetry breaking, important precision  
measurements
on the $W$, $Z$, and top quark can be made at these energies.  In this 
section, we will describe these experiments and estimate the accuracy 
they can  achieve for the realistic luminosity samples set out in  
Section~3.

To introduce this discussion, we should recall the advantageous  
features of
$\ee$ collisions that have made them so useful in the past to provide a 
detailed understanding of the underlying physics.  We will see that these
features can also be used to great advantage in the experimental program 
for 500 GeV:

\begin{figure}
\centerline{\epsfxsize=5.00truein \epsfbox{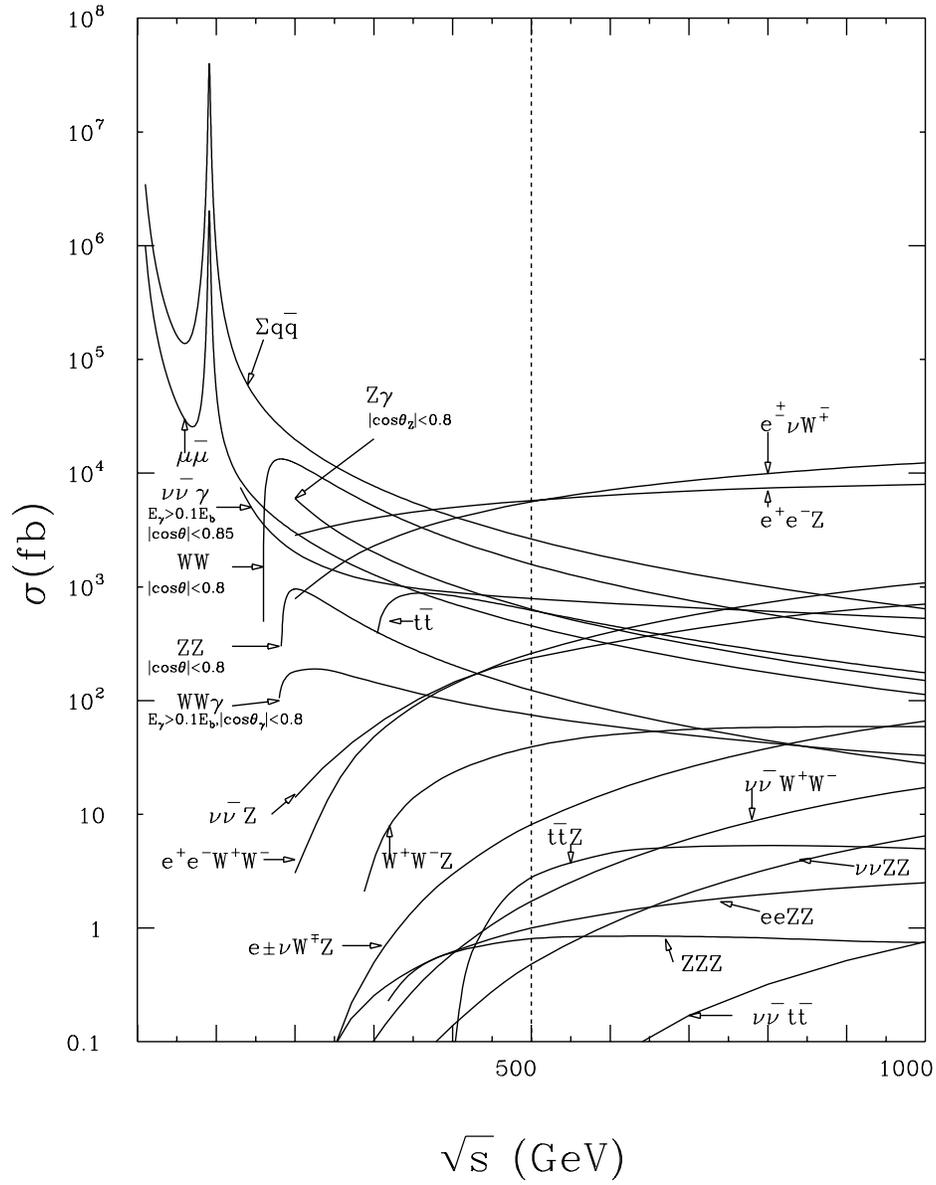}}
 \caption{\small Cross sections for a variety of physics processes  
at an $\ee$
    linear collider, from \cite{MiyamotoH}.}
\label{fig:JLCcross}
\end{figure}

\begin{itemize}
\item
The cross sections for new Standard Model and exotic  processes, and those
of the
dominant backgrounds, are all within  about 2 orders of magnitude of one
 another (see  Fig.~\ref{fig:JLCcross}).
Thus, the desired signals have large production rates and
favorable signal to background ratios.  This situation contrasts with 
that at hadron colliders,
where the interesting signals are typically very tiny fractions of  
the total
cross section.

\item
Most of the interesting processes have simple two-body
kinematics, from an initial state with  well-defined quantum numbers.

\item The cross sections for these processes are due to the electroweak
interactions and can be predicted theoretically  to part per mil accuracy.

\item
These processes also have known total energy and momentum at the  
level of the
parton-parton interaction, with
well understood and measurable
smearing from initial-state
radiation and beamstrahlung.
\item
The electron beam may be polarized, allowing
 selective suppression of
backgrounds, separation of overlapping
signals and measurement of parity-violating couplings.

\item
The collider energy may be varied
to optimize the study of particular reactions.

\end{itemize}

These features of $\ee$ collisions allow the study of heavy
 particles and their
decays in many difficult circumstances, including detection
 of decays that are
rare or have less distinct signatures, measurement of particle masses 
when some decays are invisible, measurement of spin, parity, CP, and 
electroweak quantum numbers, measurement of widths and coupling constants,
and measurement of mixing angles.

An extensive program studying physics at future high energy $\ee$  
colliders
has been carried out over the past few years as a collaborative effort of 
scientists in Europe, Asia, and America.  In this section and the next, we
will report on some highlights of that program.
Much more detail on all of these studies
can be found from  the reviews
\cite{Murayama,Accomando,SnowmassPurple,Sitges}.

\subsection{Study of the Higgs boson}

The Higgs boson plays the central role in electroweak symmetry breaking
and the generation of masses for quarks, leptons, and vector bosons.
In the Standard Model, the
Higgs boson is a simple scalar particle which couples to each fermion and
boson species proportionately to its mass.  Higher-order processes which 
couple the Higgs boson to $gg$, $\gamma\gamma$, and $\gamma Z^0$ add 
richness to its phenomenology.  If the Standard Model is not correct, 
the surprises could come at many different points.  Several scalar  
bosons could
have large vacuum expectation values and thus could share responsibility
for the $W$ and $Z$ masses. Different scalar bosons could be  
responsible for
the up- and down-quark masses, or a different boson could produce  
the masses
of third-generation fermions.  These deviations from the standard picture
might be large effects, or they might appear only in precision  
measurements.

One of the most remarkable features of the experimental environment  
of the
linear collider is its ability to probe these issues directly.  Each piece
of information---from cross sections, angular distributions, and   
branching
ratios---connects directly to a fundamental coupling of the Higgs  
particle.
In this section, we will review how measurements at a linear collider can 
assemble a complete phenomenological profile of the Higgs boson.

It is almost certain that the Higgs boson will have been discovered
before the linear collider begins operation.
Results from LEP 2 presently imply that  $m_{h} \geq$ 108 GeV
at the 95\% confidence level \cite{newLEP}.  \
It is expected that this limit will go up to
about 115 GeV as LEP 2 reaches its maximum energy.
The Tevatron may
be able to discover a Higgs boson
 up to about 180 GeV \cite{TevHiggs}.
This already covers most of the range of Higgs
boson masses favored by the arguments of Section 4.

\begin{figure}
\centerline{\epsfxsize=3.50truein \epsfbox{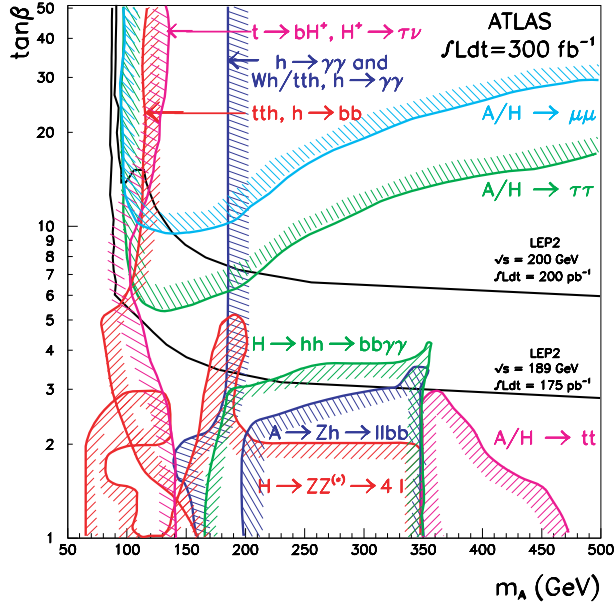}}
 \caption{\small Capability of the ATLAS experiment to study the  
Higgs sector of
         the MSSM \cite{ATLAS}.}
\label{fig:LHCHiggs}
\end{figure}

The LHC studies have shown that a Higgs boson with the
properties expected in the Standard Model can be
discovered at that facility for any value of its mass.  In addition, 
in models with an extended Higgs sector---for
example, the Minimal Supersymmetric Standard Model---the LHC should  
be able to
find one and possibly several of the Higgs particles.
  A recent summary of the
LHC sensitivity to various MSSM Higgs processes is shown in
Fig.~\ref{fig:LHCHiggs}.  There are some regions  of parameter space
for which only one channel can be observed; in any case, it
is typical that  considerable luminosity is required for positive
observation.   In Section 4.6, we have noted some specific
scenarios in which it is difficult to find the Higgs boson at the LHC.
But, more generally, the LHC is limited in its ability to assemble
a complete picture of the Higgs boson properties by the fact that
Higgs boson production is such a tiny fraction of the LHC cross
section that the Higgs particle must be reconstructed in order to
study its production and decay.                          

\subsubsection{Discovery of the Higgs independent of its decay modes}

\begin{figure}
\centerline{\epsfxsize=4.6truein \epsfbox{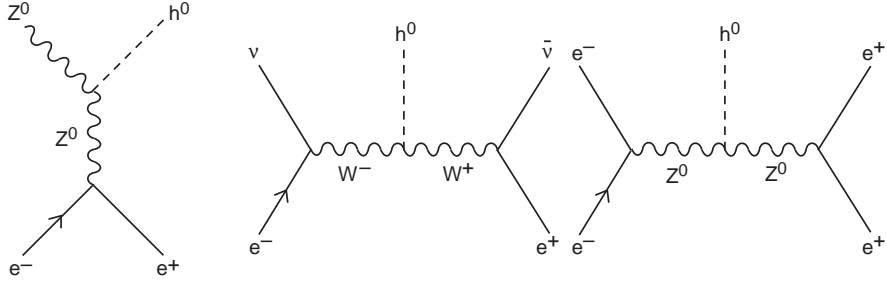}}
 \caption{\small Processes for production of the Higgs boson at an $\ee$
      linear collider.}
\label{fig:higgsproc}
\end{figure}

\begin{figure}
\centerline{\epsfxsize=4.5truein \epsfbox{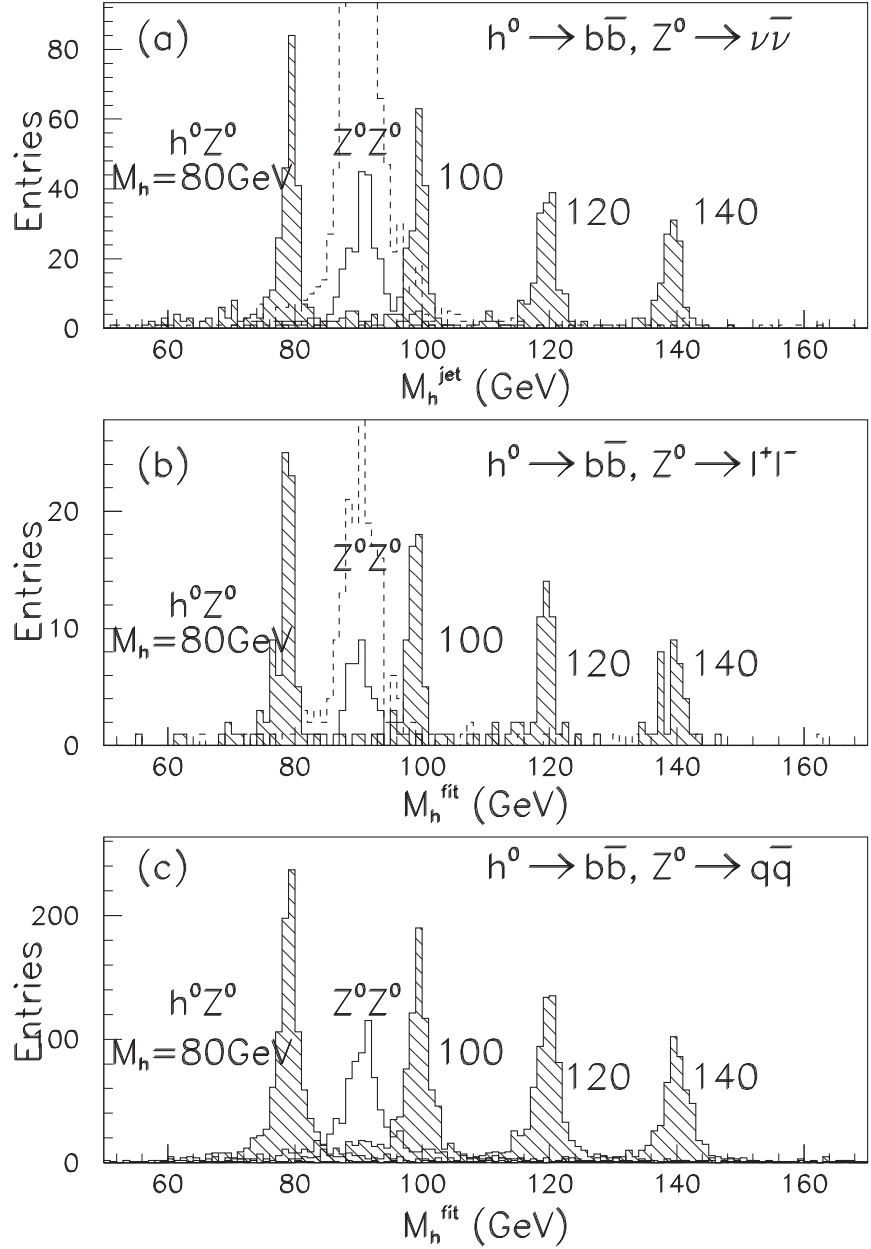}}
 \caption{\small Higgs reconstruction in the process $\ee \to Z^0  
h^0$ for
      various Higgs boson masses, using $\ell^+\ell^-$, $\nu\bar\nu$, 
       and hadronic $Z^0$ decays, for a  30 fb$^{-1}$ event sample
         at 300 GeV, from \cite{JLCone}.  The background is dominated
        by the process $\ee\to Z^0 Z^0$, which produces the
       missing-mass peak at $m_Z$.
      The unshaded solid histogram gives the background
        if a $b$-tag is applied to the Higgs candidate.  The dashed
          histograms in (a) and (b) show the background  with no   
$b$-tag.}
\label{fig:higgsdisc}
\end{figure}

As a first step, we will argue that the Higgs boson can be found at a 
linear collider whatever its decay scheme might be.  It is not
necessary to reconstruct a Higgs boson to discover the particle or to 
measure its coupling to the $Z^0$.  At low energies, the dominant
Higgs production process in $\ee$ collisions is
$\ee \to Z^{0} h^{0}$, shown as the first diagram in
Fig.~\ref{fig:higgsproc}.
If the $Z^{0}$ is
reconstructed from any one of its well-known decay modes,
 the Higgs is
seen as a peak in the missing mass distribution recoiling against  
the $Z^{0}$.
This detection is
independent of the Higgs decay mode, visible or invisible.
Simulations show that this process is very clean,  with minimal
backgrounds.  Figure~\ref{fig:higgsdisc} shows the expected signal of the 
Higgs boson using lepton, neutrino, and hadronic $Z$ decays for a
 30 fb$^{-1}$ event sample \cite{JLCone}.

The cross section for $Z^0 h^0$ production depends on the magnitude of 
the $ZZh$ coupling.  Thus, the observation of the Higgs boson in this 
process measures the size of that coupling.   If we replace the Higgs
field $h^0$ by its vacuum expectation value, we see that this same  
coupling
generates the mass of the $Z$ through the Higgs mechanism.  Thus,
determination of the
absolute magnitude of the cross section for $\ee \to Z^0 h^0$ tests
whether the observed $h^0$ generates the complete mass of the $Z^0$.
Since Higgs measurements at the LHC require reconstruction of the Higgs boson,
the LHC experiments can only measure ratios of couplings and
cannot determine the $ZZh$ coupling directly.

 If there are several
Higgs bosons contributing to the mass of the $Z^0$, the $\ee$ cross  
section
for production of the lightest Higgs will be smaller, but heavier Higgs
bosons must appear at higher values of the recoil mass.  To discuss this 
quantitatively, let the coupling
of the boson $h_i$ be $g_{ZZi}$. (For simplicity, we assume that all
of the $h_i$ are $SU(2)$ doublets; this assumption can be checked by  
searching
 for multiply-charged Higgs states.)  Then the statement that the sum 
of the contributions from the vacuum expectation values of the $h_i$
generates the full mass of the $Z^0$ can be expressed as the sum
rule \cite{GunionHaber}
\beq
      \sum_i g_{ZZi}^2  =    4 {\mz^4}/v^2   \ ,
\eeq{GHsumrule}
where $v = 246$ GeV.  With a 200 fb$^{-1}$ event sample at 500 GeV,
Higgs particles $h_i$ can be discovered in recoil against the $Z^0$
down to a cross section of 0.2 of the Standard Model value  for
$m(h_i) = 350$ GeV, and below 0.01 of the Standard Model value for
$m(h_i) = 150$ GeV \cite{SnowmassPurple}.  If all contributing
Higgs bosons have masses
below 150 GeV, the sum rule can be checked in a 200 fb$^{-1}$ experiment
to  5\% accuracy, with dominantly statistical uncertainty.
  When we have  
saturated the
 sum rule \leqn{GHsumrule}, we will have discovered all of the Higgs  
states
that contribute to the $Z^0$ mass.

\subsubsection{Measurement of the Higgs branching ratios}

The Higgs boson branching ratios are crucial indicators of 
nature of this particle, and of possible extensions beyond the
Standard Model.  The LHC can only make rough measurements of these, to
about the 25\% level,
and only for some values of the Higgs boson mass \cite{ATLAS,Rainwater}.
Once the mass is known, it is straightforward at the linear collider to
measure Higgs boson absolute branching fractions
into two fermion or two gauge bosons for any of the
production processes of Fig.~\ref{fig:higgsproc}
using the energy and momentum constraints.
All decay modes of the $Z^0$ can be used
in this study,
even $Z^0 \to \nu\bar\nu$ (20\% of the $Z^0$ total width) \cite{Battaglia}.

\begin{figure}
\centerline{\epsfxsize=6.00truein \epsfbox{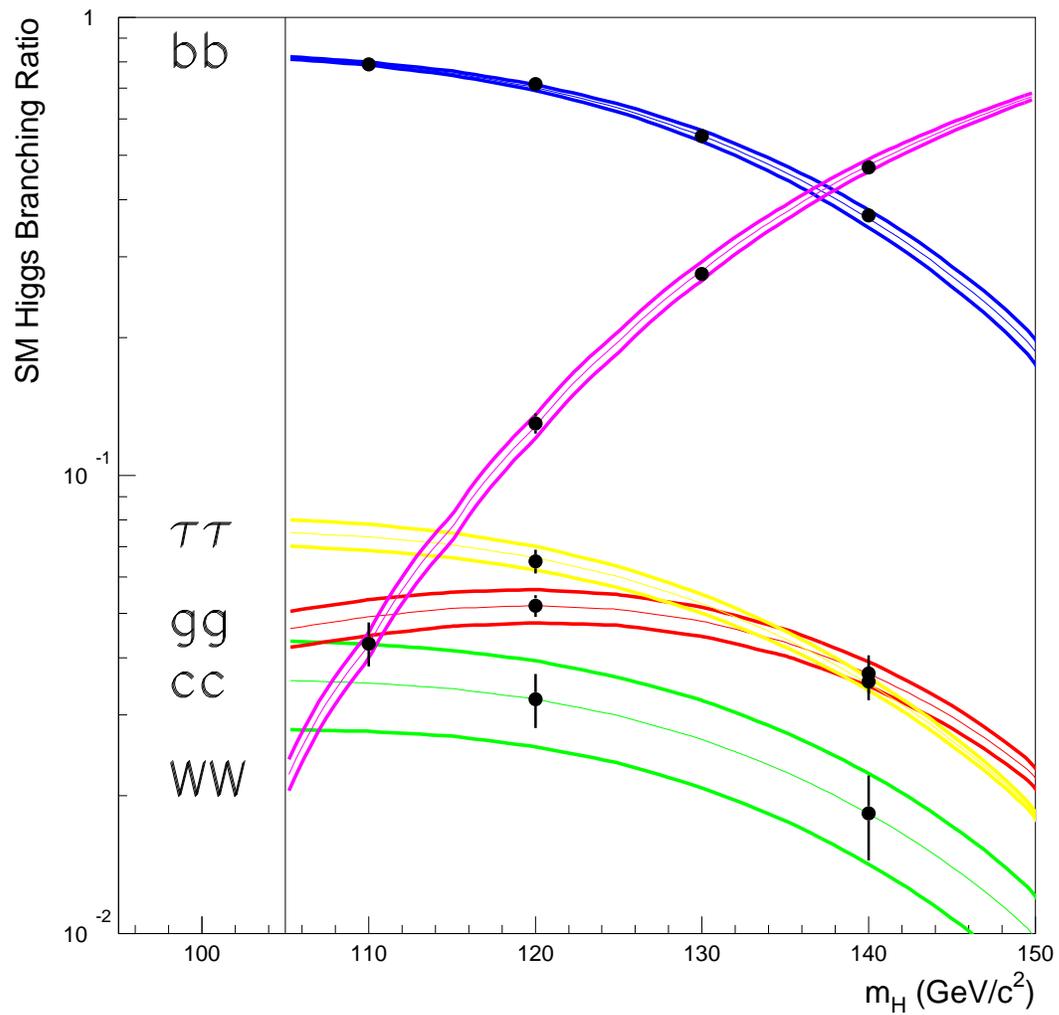}}
 \caption{\small Determination of Higgs boson branching ratios in a  
variety of
        decay modes, from \cite{Battaglia}.  The error bars show the 
        expected experimental errors for 500 fb$^{-1}$ at 350 GeV.  The
         bands show the theoretical errors in the Standard Model  
predictions.}
\label{fig:Battaglia}
\end{figure}

 Methods for determining the Higgs cross sections to various
decay channels have been studied recently in \cite{Battaglia}.  It is 
straightforward that the $b\bar b$ decays can be identified by vertex 
tagging.  The studies show that $c\bar c$ decays can also be
identified by vertex tagging with high efficiency, since the first
layer of a vertex detector can be placed at about  1 cm from the  
interaction
point.   Multi-jet decays of the $h^0$ are  typically $WW^*$.
Table~3 gives a summary of the
precision
expected for a large variety of decay modes for the case of a 120 GeV
Higgs boson.  This case is especially favorable in terms of the number
of final states which are accessible, but it is also the value of the 
Higgs mass which is most probable in the Minimal Supersymmetric  
Standard Model.
Expectations for Higgs branching ratio measurements at other values  
of the
Higgs mass (assuming 500 fb$^{-1}$ at 350 GeV) are shown in
Fig.~\ref{fig:Battaglia} \cite{Battaglia}.    
If the Standard Model Higgs mass approaches 200 GeV, the dominance of
the $WW$ and $ZZ$ decays will render the fermionic decays progressively more
difficult to observe.

\begin{table}
\begin{center}
\begin{tabular}{|cc|r|r|}\hline
     &      &   200 fb$^{-1}$ &  500 fb$^{-1}$\\[.5ex] \hline\hline
${\Delta\sigma_{ZH}/\sigma_{ZH}}$& &  4\%  & 3\% \\[.5ex] \hline\hline
\raisebox{-.25ex}
{$\Delta\sigma_{H\nu\nu}BR(b\bar b)/\sigma_{H\nu\nu}BR(b\bar b)$} &&
\raisebox{-.25ex}{ 3\%}& 2\% \\[.5ex] \hline\hline
\raisebox{-.25ex}{${\Delta BR/BR}$}
& \raisebox{-.25ex}{ $b\bar b$}
& \raisebox{-.25ex}{ 3\%} &\raisebox{-.25ex}{ 2\%} \\[.5ex]
& $WW^*$ &  8\% & 5\% \\
& $\tau^+\tau^-$ &  7\% & 6\%\\
& $c\bar c$ &  10\%  & 8\% \\
& $gg$ &  8\% & 6\% \\
& $\gamma\gamma$ &  22\% & 14\% \\[.5ex] \hline
\end{tabular}
\caption{\small Expected errors in branching ratio and coupling  
measurements
for a Standard Model Higgs boson of mass 120 GeV, from measurements
at 350 GeV.}
\end{center}
\label{tab:HiggsBRs}
\end{table}

The Higgs branching ratios directly address the question of whether the 
Higgs boson generates the masses of all Standard Model particles.  If
the vacuum expectation value of $h^0$ produces the fermion masses,
the couplings of $h^0$ to $b$, $c$, and $\tau$ should be simply 
determined from the 
ratio of their masses.  Similarly, the coupling of the $h^0$ to $WW$ or, 
for the case of a light Higgs, to one on-shell and one off-shell $W$, 
measures the fraction of the $W$ mass due to the Higgs vacuum expectation
value.

The Minimal Supersymmetric Standard Model includes an extended Higgs  
sector
with two $SU(2)$ doublets.  For the most general case of a  
two-Higgs-doublet
model, vacuum expectation values of both Higgs fields contribute to the 
quark, lepton, and boson masses and the predictions for branching ratios 
differ qualitatively from those in the Standard Model.  However, in the 
MSSM with heavy superpartners, one scalar boson $H^0$ is typically heavy
and the orthogonal boson $h^0$, which must be light, tends to  resemble
 the Higgs boson of the Standard Model.  For example, the ratio of
branching ratios to $b\bar b$ and $WW^*$ is corrected by the factor
\beq
       1 +  2 \cos^2 2\beta \sin^2 2\beta \, {\mz^2\over m_H^2} +  
\cdots \ .
\eeq{MSSMcorrection}
Nevertheless, accurate branching ratio measurements can distinguish the 
MSSM Higgs boson from the Standard Model Higgs boson over a large region
of parameter space.  From the results of \cite{Battaglia}, the 500  
fb$^{-1}$ 
experiment discussed above would exclude corrections from the MSSM Higgs
structure for $m_A$ up to at least 
550 GeV.  The linear collider determination of 
branching ratios is sufficiently accurate that the theoretical  
uncertainty
in the charm quark mass is  actually the dominant source of error.  New
approaches to the determination of the quark masses in lattice gauge  
theory
should give more accurate values in the next few years \cite{Kronfeld}
and thus improve the power of this measurement.

\subsubsection{Measurement of the Higgs boson width}

It will be critical to know the total width of the Higgs, 
$\Gamma_{\rm tot}$, accurately.
For a Higgs boson mass below 200 GeV, the total width is expected to be 
below 1 GeV, too small to be measured at the LHC or 
directly at the linear collider.  
To determine this
width, one will need to combine an absolute measurement of a decay
rate or coupling constant with the measurement of the branching ratio for 
the corresponding channel.  The most promising method is to use the
branching ratio to $WW^*$.  The absolute size of the $WWh$ coupling can be
determined either from the $SU(2)\times U(1)$ relation
 $g^2_{WWh}/g^2_{ZZh} = \cstw$ or, in a more model-independent
way, from the cross section for $h^0$ production by the $WW$ fusion  
process
shown as the second diagram in Fig.~\ref{fig:higgsproc}.  (The $ZZ$ fusion
process is expected to add only  a small contribution.)
From Table~3, the Higgs branching
ratio to $WW^*$ gives the dominant source of error in this measurement.

If the $\gamma\gamma$ collider option is realized by backscattering  
polarized
laser light off the $e^{\pm}$ beams, then the process $\gamma\gamma
\rightarrow h^{0}$ can be used to measure the absolute partial width
$\Gamma(h^0\to \gamma\gamma)$.  This width, which can be determined  
to about
5\% accuracy with a 200 fb$^{-1}$ dedicated experiment \cite{gamgamH},
is of great interest
in its own right, since it measures a sum of contributions from all
heavy charged  particles that couple to the $h^0$.

\subsubsection{Measurement of the spin-parity and CP of the Higgs boson}

It will be essential to determine the quantum numbers of an
observed Higgs boson unambiguously.  The LHC can rule out spin 1
 if the decay $H\rightarrow\gamma\gamma$ is observed.  If the
decay  $H\rightarrow Z Z$ is observed, spin 0 and 1 could be distinguished at
the LHC, but the CP quantum numbers will be difficult to determine in any case.
The linear collider will thus be needed to determine the 
Higgs quantum numbers.

\begin{figure}
\centerline{\epsfxsize=3.00truein \epsfbox{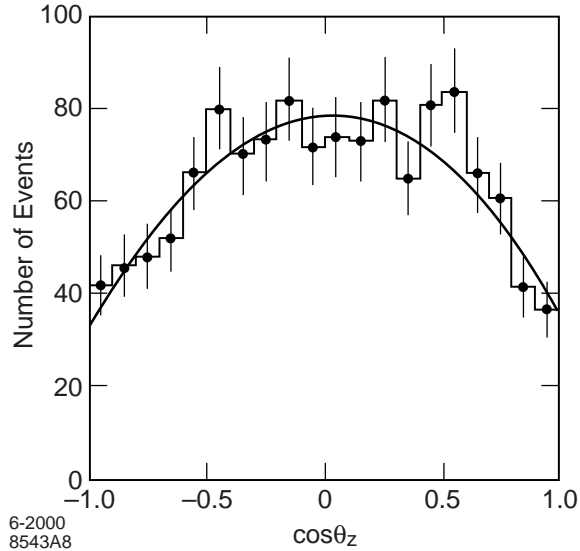}}
 \caption{\small Angular distribution of the $Z$ boson in $\ee\to Z^0 h^0$, as
  reconstructed from a 50 fb$^{-1}$ event sample at 300 GeV, from
   \cite{JanotHawaii}.}
\label{fig:AD}
\end{figure}

If the Higgs field has a vacuum expectation value, it must be a CP-even
spin-0 field.  Thus, a Higgs boson produced in $\ee \to Z^0 h^0$
with a rate comparable to the
Standard Model rate must have these quantum numbers.  However, there are
a number of checks on these properties that are available from the
kinematics of Higgs production.  In the limit $s \gg \mz^2, \mh^2$,  
a scalar
 Higgs boson produced in this reaction has an angular distribution
\beq
   {d\sigma\over d\cos\theta} \sim  \sin^2\theta \quad ,
\eeq{Higgsangles}
and the $Z^0$ recoiling against it is dominantly longitudinally polarized,
and so that distribution in the decay angle peaks at central values. 
(For a CP-odd scalar, these distributions differ qualitatively.)
If the center of mass energy is not asymptotic, the corrections to these
relations are predicted from kinematics.  For example, Fig.~\ref{fig:AD}
shows a simulation of the angular distribution at
300 GeV and a comparison to the distribution expected for a Higgs scalar.

The production of the Higgs boson in $\gamma\gamma$ collisions goes  
through
a loop diagram which can give both scalar and pseudoscalar couplings.
Thus, the $\gamma\gamma$ collider option offers a nontrivial test of 
CP violation.  With longitudinal $\gamma$ polarization, the asymmetry
of Higgs production cross sections
\beq
   A_{\gamma} =  {\sigma(\gamma_L\gamma_L) -  
\sigma(\gamma_R\gamma_R)\over
\sigma(\gamma_L\gamma_L) +\sigma(\gamma_R\gamma_R)}
\eeq{CPsignal}
vanishes for pure scalar or pseudoscalar coupling to $\gamma\gamma$ but is
nonzero if the Higgs is a mixture of CP eigenstates.  Models with CP  
violation
in the top sector can give 10\% or larger asymmetries \cite{GrzG}.
In models with extended Higgs sectors, this polarization asymmetry can 
incisively separate the heavy scalar and pseudoscalar Higgs resonances
\cite{Asakawa}.

\subsubsection{Measurement of the Higgs self couplings}

The Higgs self-couplings are uniquely fixed in the Standard Model
in terms of the Higgs field expectation value $v$; in the minimal
supersymmetric model, they depend on the Higgs field couplings and mixings.
Measuring the self-couplings is a crucial step in checking the consistency 
these models, and it  gives added information on the parameters of
supersymmetric models.  It appears that observation  of Higgs pair
production at the LHC will be very difficult due to the dominance  
of gluon
fusion production and large QCD backgrounds \cite{DjouadiLHC}.
In $e^+e^-$ collisions, production of two Higgs bosons
in the final state can occur for any of the diagrams of
Fig.~\ref{fig:higgsdisc} by radiating an additional Higgs from any of the 
gauge
boson legs, or through the trilinear Higgs coupling.  The cross sections for
production of a pair of Higgs bosons
  with an associated $Z$ boson have been calculated
 to be of order 0.5 fb for $\mh = 110$ GeV at $\sqrt  
s = 500$ GeV
in the Standard Model \cite{DjouadiLC}.   Cross sections for various 
supersymmetric  Higgs pair-production processes are comparable for 
much of the supersymmetric parameter  
space.  The
final state of $Z h h$, with both Higgs bosons observed as $b\bar b$,
 should provide a detectable
signature without large backgrounds, yielding a precision on the trilinear
Higgs coupling of roughly 25\% for 600 fb$^{-1}$.

\subsection{Studies of supersymmetry}

In Section 4, we argued that the new physics at the TeV energy scale is 
likely to be a supersymmetric extension of the Standard Model.  If
supersymmetric particles appear at the next step in energy, they will
provide a rich field for experimental study.  This study will address two
separate and important issues.  First, supersymmetry entails a fundamental
modification of the structure of space-time.  Supersymmetry can be  
described
as the statement that spinors and fermions are an integral part of  
space-time
geometry, or, alternatively, that there are new space-time  
dimensions which
are fermionic in character. It requires new gravitational equations that 
include a spin-$\thalf$ partner of the graviton.  If we are to claim that 
Nature has this structure, we must to prove it
experimentally by demonstrating the quantum number assignments and  
symmetry
relations that this structure requires.

Second, phenomenological models with supersymmetry introduce a large  
number
of new physical parameters. The masses of supersymmetric particles,  
and other
parameters associated with spontaneous supersymmetry breaking, are
not fixed from currently known principles
but, rather, must be determined experimentally.
The most general description of supersymmetry breaking even in the  
`Minimal'
Supersymmetric Standard Model contains 105
parameters. Each  explicit model of spontaneous supersymmetry  
breaking gives
predictions for these parameters or relations among them.
  But there is no `Standard Model' of supersymmetry
breaking. In the literature, one finds at least three general
approaches---gravity-,
gauge-, and anomaly-mediation---each of which has numerous variants. 
Each approach is derived from assumptions about new physics at a  higher
energy scale, which ranges from $10^5$ to $10^{19}$ GeV depending on the 
model.    
The various models predict
mass spectra and mixing parameters that  differ characteristically. These
observables 
provide clues to the nature of physics at extremely short distances, 
possibly even to the truly fundamental physics at the scale of grand 
unification or quantum gravity \cite{EISSB}.

Supersymmetric particles may well be discovered in Run II of the  
Tevatron.       In any case, if supersymmetry is relevant to 
    electroweak symmetry breaking, 
supersymmetric particles should surely be found at the LHC.  
The LHC
collaborations have demonstrated that they would be  sensitive to quark
and gluon superpartners up to masses of at least 2 TeV.  For the gluino, this
reach goes about a factor of 2 beyond the fine-tuning limits given in 
Table~2.  Reactions which produce the squarks and gluinos
also produce the lighter supersymmetric particles into which they decay.
The ATLAS and CMS collaborations have presented some striking analyses at
specific points in the parameter space of mSUGRA models in which 3 to 5 
mass parameters can be determined from kinematics.  From this information,
the four parameters of the mSUGRA model can be determined to 2--10\%
accuracy \cite{ATLAS,CMS}.

Ultimately, though, hadron colliders are limited in their ability to probe
the underlying parameters of supersymmetric models. 
Because the LHC  produces many SUSY particles and observes
many of their decay chains simultaneously,
it is  difficult  
to isolate
parameters and determine them in a model-independent way.  It is difficult
to determine the spin and electroweak quantum numbers of particles
unambiguously.  And, only limited information can be obtained about the 
heavier color-singlet particles, including sleptons and  heavier  
charginos and
neutralinos, and about the unobserved lightest neutralino.

It is just for these reasons that one needs a facility that can  
approach the
spectroscopy of supersymmetric particles from an orthogonal direction. 
An $\ee$ collider can study supersymmetric particles one at a time,
beginning with the lightest and working upward to particles with  
more complex
decay patterns.  For each particle, the measurements go well beyond
simple mass determinations.  We will give a number of illustrative  
examples
in this section.

To carry out  these measurements,
 it is only necessary that supersymmetric particles
can be pair-produced at
the energy provided by the $\ee$ collider.  In the program that we have
presented in Section 2, in which a collider with an  initial energy of
500 GeV evolves to higher center of mass energies, one 
can eventually create the full set of supersymmetry particles.
Here we concentrate on 
the expectations for 500 GeV.  In  Section 4.4, we have argued
that the lightest charginos and neutralinos, the supersymmetric partners 
of the photon, $W$, $Z$, and Higgs bosons,
should be produced already at the initial 500 GeV stage.
 The mSUGRA models
discussed in Section 4.4 do not place such strong constraints on the 
masses of lepton superpartners, but in other schemes of supersymmetry
breaking, such as gauge-mediation and the no-scale limit of  
gravity-mediation,
it is natural for the sleptons to be as light as the charginos.  Because
the experimental study of sleptons is conceptually very simple, we  
will present
the linear collider experimental program for sleptons in this  
section along
with our discussion of charginos.  Other issues for the experimental 
study of supersymmetry will be discussed in Section 6.2.

Our discussion of the basic supersymmetry measurements in this  
section will
be rather detailed. In reading it, one should keep in mind that
the linear collider
offers a similar level of detailed information for any other new particles
that might appear in its energy range.

\subsubsection{Slepton mass measurement}

The simple kinematics of supersymmetric particle pair production allows
direct and accurate mass measurements.  The technique may be illustrated
with the process of pair production and decay of the ${\s\mu}_R^-$, the scalar
partner of the $\mu^-_R$. The process
 $\ee\to {\s\mu}_R^- {\s\mu}_R^+$
produces the sleptons at a fixed energy equal to the beam energy.
 The $\s\mu_R^-$ is expected to decay to the unobserved lightest
neutralino via ${\s\mu}^-_R \to \mu^- \neu1$.  Then the final muons
are distributed in energy between kinematic endpoints determined by
the masses in the problem.    Since the ${\s\mu}^-_R$
is a scalar, the distribution of muons is isotropic in the  ${\s\mu}^-_R$
rest frame and flat in energy in the lab frame.  Thus, the observed
energy distribution of muons has the shape of a rectangular box, and the  
masses of both  the ${\s\mu}^-_R$ and the $\neu1$ can be read off  
from the
positions of the edges.

In measuring slepton pair production
in $e^+e^-$ collisions, special attention 
 must be paid to the backgrounds from two-photon  
processes in which the primary scattered electrons are undetected within
the beam pipes.  This makes it important for the detector to have good 
coverage at  forward and backward angles. It may be useful for gaining 
further control over this process to provide tagging detectors at very  small 
angles \cite{Danielson}.

\begin{figure}
\centerline{\epsfxsize=7.00truein \epsfbox{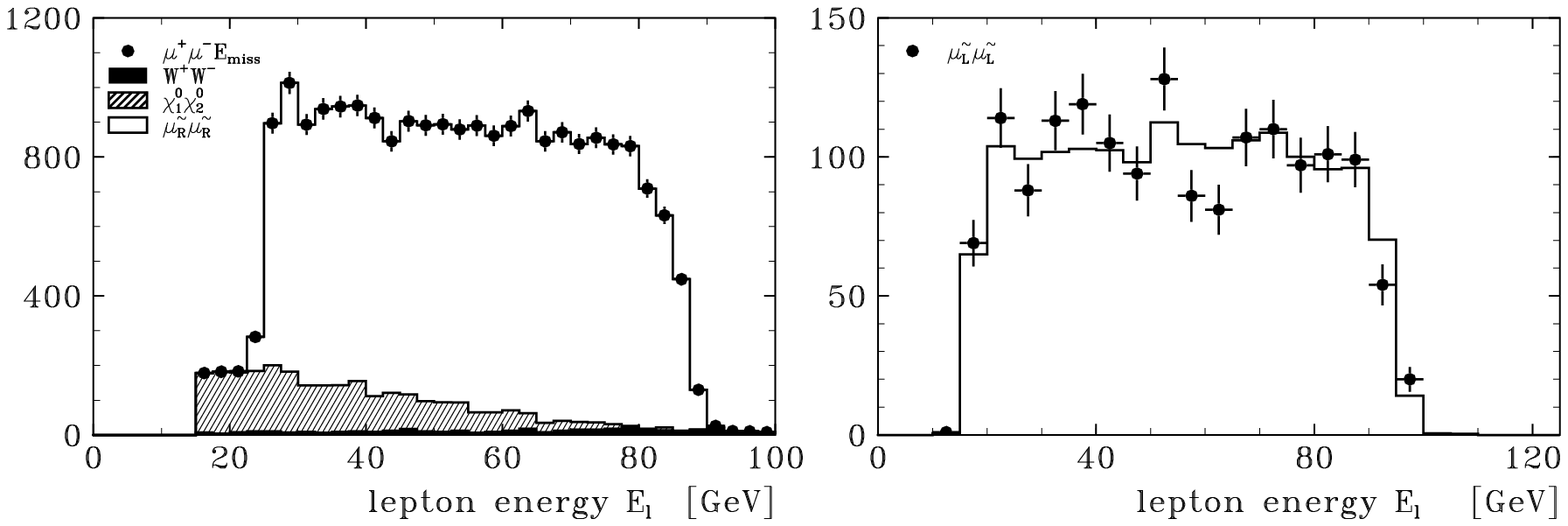}}
 \caption{\small Energy distribution of muons resulting from processes
    $\ee\to  {\s\mu}^- {\s\mu}^+$, followed by $\s\mu$ decay,
  from \cite{MBSUSY}.
  left:  $\ee\to  {\s\mu}_R^- {\s\mu}_R^+$, for a 160 fb$^{-1}$ event
   sample at 320 GeV; right:  $\ee\to  {\s\mu}_L^- {\s\mu}_L^+$,
       with selection of ${\s\mu}_L\to \mu \neu2$, 
$\neu2 \to \ell^+\ell^-\neu1$  decays on both sides, 
 for  a  250 fb$^{-1}$ event
   sample at 500 GeV.
 The electron beam polarization is used to reduce the background
     from $\ee\to W^+W^-$.}
\label{fig:MB}
\end{figure}

On the left side of Fig.~\ref{fig:MB}, we show simulation results for 
${\s\mu}_R$ pair production \cite{MBSUSY}.  The dominant background  
(shaded
in the figure) comes from other supersymmetry processes.  The rounding of 
the rectangle on its upper edge is the effect of beamstrahlung and
initial state radiation.  The simulation predicts a measurement of both
the  ${\s\mu}^-_R$ and the $\neu1$ masses to 0.2\% accuracy.  The right
side of  Fig.~\ref{fig:MB} shows the muon energy distribution from  pair
production of the   ${\s\mu}^-_L$, the partner of the $\mu^-_L$.   
Decays of
the form  ${\s\mu}_L \to \mu \neu2$, $\neu2 \to \ell^+\ell^-\neu1$ are
selected on both sides of the event to obtain a very clean 6 lepton
signature.  Despite the low statistics from the severe event selection,
this analysis also gives the  ${\s\mu}^-_L$ and the $\neu2$ masses to 
0.2\% accuracy. At the LHC, the mass of the lightest neutralino $\neu1$
typically cannot be
determined directly, and the masses of heavier superparticles 
are determined
 relative to
the $\neu1$ mass. 
So not only do the $\ee$  measurements provide
model-independent slepton masses, they also provide crucial  
information to
make the superpartner mass measurements from the LHC more  
model-independent.

The same strategy can be applied to determine the masses of
other superpartners.  Examples with sneutrinos, scalar top, and charginos
are shown in \cite{SnowmassSUSY}.  Even higher accuracies
can be obtained by scanning the $\ee$ cross section
near each pair production threshold.  This costs about 100 fb$^{-1}$ per 
threshold, but it allows particle mass measurements to better than
1 part per mil \cite{MBSUSY}.

\subsubsection{Slepton properties}

An $\ee$ collider can not only measure the masses of superparticles  
but also can 
determine many more properties of these particles, testing predictions 
 of supersymmetry from the most
qualitative to the most detailed.

Before anything else, it is important to verify that particles that  
seem to
be sleptons are spin 0 particles with the Standard Model quantum  
numbers of
leptons.  A spin 0 particle has a characteristic angular distribution in 
$\ee$ annihilation, proportional to $\sin^2\theta$.  Even though  
there are
missing neutralinos in the final state of
  $\ee\to  {\s\mu}^- {\s\mu}^+$, there are enough kinematic  
constraints that
the angular distribution can be reconstructed \cite{Tsukamoto}.
The magnitude of the cross section can be computed for each electron 
polarization with typical electroweak precision; it depends only on the 
Standard Model quantum numbers of the produced particle and thus  
determines
these quantum numbers.

A major issue in supersymmetry is the flavor-dependence of supersymmetry
breaking parameters.  Using the endpoint technique above, the  
selectron and
smuon masses can be compared at a level below the 1 part per mil level.  
It is somewhat more difficult to study the superpartners of the  
$\tau$, but
even in this case the masses can be found to percent accuracy by  
locating the
endpoint of the energy distribution of stau decay products \cite{FNT}.

\begin{figure}
\centerline{\epsfxsize=3.00truein \epsfbox{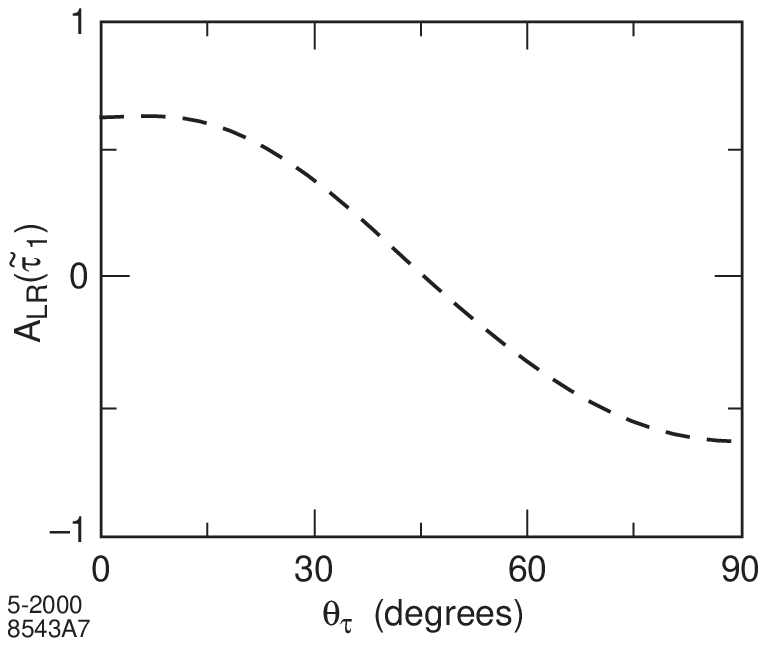}}
 \caption{\small Polarization asymmetry of $\ee \to \s \tau_1^+  \s  
\tau_1^-$
   as a function of the stau mixing angle.}
\label{fig:tauasym}
\end{figure}

It is typical in supersymmetry scenarios with large $\tan\beta$ that the 
superpartners of $\tau_R^-$ and $\tau_L^-$ mix, and that the lighter mass
eigenstate is actually the lightest slepton.  If the mass difference  
between
the lighter stau and the other leptons is significant, this can create a
problem for the study of supersymmetry at LHC, since then  
supersymmetry decay
cascades typically end with $\tau$ production.  A parameter
point studied by the
ATLAS supersymmetry group illustrates the problem \cite{ATLAS}.
We have just noted that there is no difficulty  in measuring the  
stau masses
at a linear collider.  In addition, since
 the production cross section depends only on
electroweak quantum numbers,
it is possible to determine the mixing angle from total cross section and
polarization asymmetry measurements.  The characteristic dependence of the
polarization asymmetry on the stau mixing angle is shown in
Fig.~\ref{fig:tauasym}.  The final state $\tau$ polarization  
provides another
diagnostic observable which can be used to analyze the composition of the 
stau or of the neutralino into which it decays \cite{FNT}.

\begin{figure}
\centerline{\epsfxsize=3.00truein \epsfbox{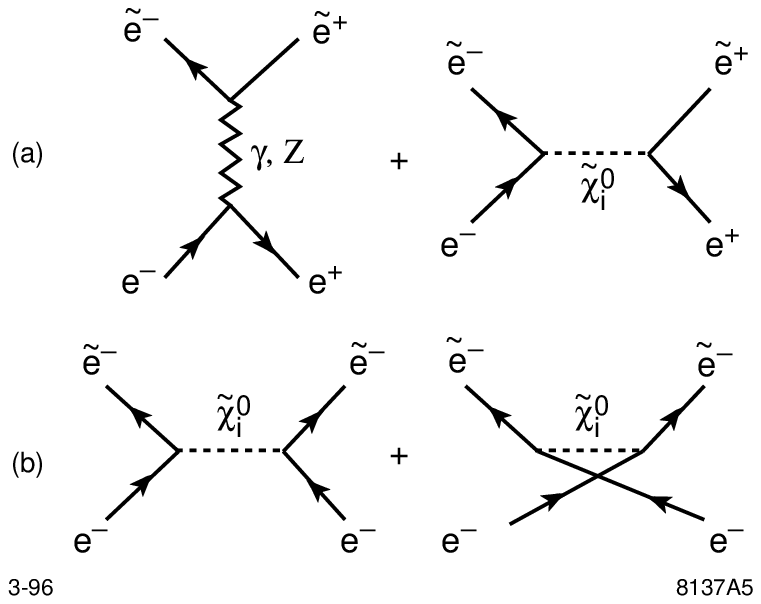}}
 \caption{\small Diagrams contributing to selectron pair production: 
(a) $\ee \to {\s e}^+ {\s e}^-$, (b) $e^-e^- \to {\s e}^- {\s e}^-$.}
\label{fig:selectronprocs}
\end{figure}

The cross section for production of the electron partners is somewhat
more complicated, because this process can proceed both by $\ee$
annihilation and by the exchange of neutralinos, as shown in
Fig.~\ref{fig:selectronprocs}.  In typical models, the
dominant contribution actually comes from exchange of the lightest
neutralino.  Thus, the selectron production cross section can give further
information on the mass and the properties of this particle.  The study of
neutralinos is complicated by the fact that the various neutralino species
can mix.  In the Section 5.2.4, we will discuss this mixing problem and 
present methods for resolving it experimentally using $\ee$ data on
chargino production.  Neutralino mixing can also be studied in
selectron pair production; an illustrative analysis is given in
\cite{Tsukamoto}.

\begin{figure}
\centerline{\epsfxsize=3.00truein \epsfbox{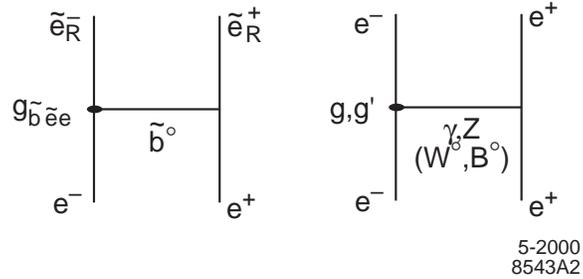}}
 \caption{\small Comparison of the weak interaction coupling  
$g^\prime$ and its
     supersymmetric counterpart $g_{\tilde{b}\tilde{e}_R e}$.}
\label{fig:ginee}
\end{figure}

Once the mixing of neutralinos is understood, the selectron pair
production can test the basic idea of supersymmetry quantitatively,  
by testing
the symmetry relation of coupling constants.  For simplicity, consider a 
model in which the lightest neutralino is the superpartner $\s b$
of the $U(1)$
gauge boson of the Standard Model, and imagine comparing the processes of
${\s e}_R$ pair production and Bhabha scattering, as
illustrated in Fig.~\ref{fig:ginee}.  By supersymmetry, the coupling  
constant
at the $e \s e \s b$ vertex must be simply related to the $U(1)$  
electroweak
coupling: $g_{\tilde{b}\tilde{e}_R e} = \sqrt{2} g'$. A measurement of the
forward
cross section for  $\ee \to {\s e}_R^+ {\s e}_R^-$ can give a precision 
test of this prediction.

\begin{figure}
\centerline{\epsfxsize=3.00truein \epsfbox{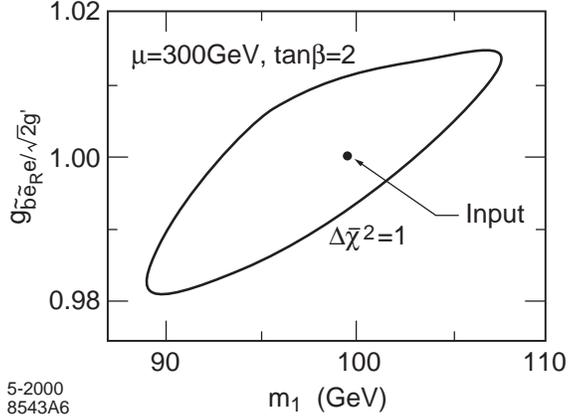}}
 \caption{\small Determination of the  $g_{\tilde{b}\tilde{e}_R e}$
coupling from a 100 fb$^{-1}$
measurement of selectron pair production, from \cite{FNT}.}
\label{fig:Nojiri}
\end{figure}

Detailed simulation of selectron pair production
has shown  that the ratio
$g_{\tilde{b}\tilde{e}e}/\sqrt{2}g^\prime$ can be measured to a precision 
of about 1\%, as shown in Fig.~\ref{fig:Nojiri} \cite{FNT}.
(This analysis uses data from the same cross section measurement both to
 fix the parameters of the neutralino mixing and to determine
$g_{\tilde{b}\tilde{e}e}$.)
 Even higher accuracy can be achieved by studying selectron
production in $e^-e^-$ collisions.  The ratio $g_{\tilde{W}\tilde{\nu}e}$
can also be determined from chargino pair production and compared to its
Standard Model counterpart to about 2\% accuracy.
At these levels, the measurement would not only provide a stringent
test of supersymmetry as a symmetry of Nature, but also it
might be sensitive to radiative corrections from heavy squark  
and slepton
species \cite{FengMC,NojiriPY,RandallS}.

\subsubsection{Chargino mass measurement}

\begin{figure}
\centerline{\epsfxsize=5.00truein \epsfbox{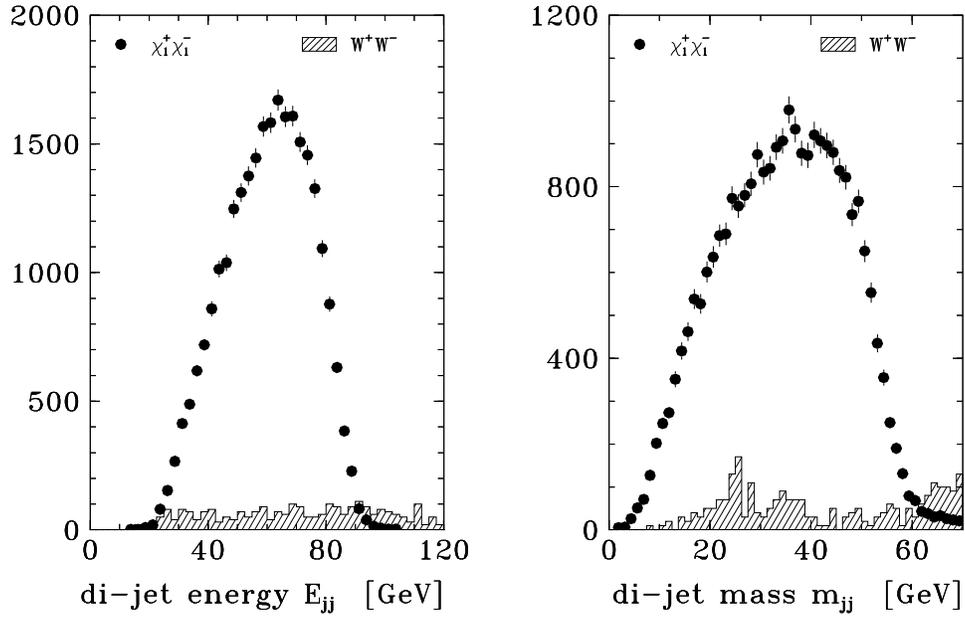}}
 \caption{\small Kinematic distributions from a simulation of  
chargino pair
production and decay with 160 fb$^{-1}$ at 320 GeV,
 from \cite{MBSUSY}. left: dijet energy distribution;
right: dijet mass distribution.}
\label{fig:charginomass}
\end{figure}

The process of chargino pair production in $\ee$ annihilation is somewhat
more complicated than slepton pair production, but it also provides more 
interesting observables.  To begin, we discuss the chargino mass
measurement.  If the chargino is the lightest charged supersymmetric  
particle,
it will decay via $\ch1 \to q\bar q \neu1$ or $\ch1 \to \ell^+ \nu \neu1$.
The reaction with a hadronic decay on one side and a leptonic decay
on the other
provides a characteristic sample of events which can be  
distinguished from $W$
pair production by their large missing energy and momentum.  If the lab 
frame energy of the $q\bar q$ system is measured, the kinematic  
endpoints of
this distribution can be used to determine the mass of the $\ch1$ and of 
the $\neu1$, as in the slepton case.
The power of this kinematic fit can be
strengthened
by segregrating events according to
the measured value of the $q\bar q $ invariant mass.  
 The distributions in 
 the energy and mass of the $q\bar q$  system are shown
  in Fig.~\ref{fig:charginomass}. 
In the study of 
\cite{MBSUSY}, one finds mass determinations at the 0.2\% level for 
event samples of the same size as those used in the slepton case.

At large $\tan\beta$ values, the lighter stau ($\tilde{\tau}_{1}$)
may be lighter than the lightest chargino ($\tilde{\chi}^{\pm}_{1}$).
The decay $\tilde{\chi}^{\pm}_{1} \rightarrow \tilde{\tau}^{\pm}_{1}
\nu_{\tau}$, followed by $\tilde{\tau}^{\pm}_{1} \rightarrow$
$\tilde{\chi}^{0}_{1}\ \tau^\pm$, alters the phenomenology of the chargino
production \cite{FNT}. In this case, one can still measure
the mass of a 170 GeV chargino to better than 5 GeV with 200 fb$^{-1}$ at
$\sqrt{s}$ = 400 GeV \cite{Kamon}.

\begin{figure}
\centerline{\epsfxsize=2.50truein \epsfbox{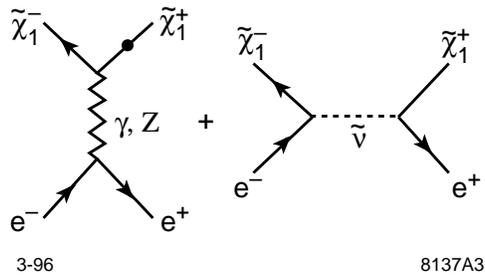}}
 \caption{\small Diagrams contributing to chargino pair production.}
\label{fig:chdiags}
\end{figure}

\subsubsection{Analysis of chargino mixing}

The cross section and angular distribution of chargino pair production is
built up from the diagrams shown in Fig.~\ref{fig:chdiags}.  This
process is intrinsically more complicated than slepton pair production
because one must account for chargino mixing.  In supersymmetry models, 
there is always a charged Higgs boson $H^\pm$, and both the $W^\pm$  
and the
$H^\pm$ have spin-$\half$ partners.  These necessarily mix, through  
a mass
matrix of the following form:
\beq
               \pmatrix{\s w^- & i\s h^-_1\cr}^T
 \pmatrix{ m_2 &   \sqrt{2}\mw \sin\beta \cr
                    \sqrt{2}\mw \cos\beta & \mu\cr }
 \pmatrix{ \s w^+ \cr i \s h^+_2\cr}\ ,
\eeq{chmatrix}
where $\s w^\pm$ are the superpartners of the $W^\pm$ and $\s h^-_1$ and
$\s h^+_2$ are the superpartners of the charged components of the two
Higgs fields.  The matrix depends on the parameters $\mu$, the  
supersymmetric
Higgs mass, $m_2$; the supersymmetry breaking mass of the $\s w^\pm$; and
$\tan\beta$, the ratio of Higgs field vacuum expectation values.
The neutralino masses involve a similar mixing problem among four states,
the superpartners of the neutral $SU(2)$ and $U(1)$ gauge bosons and  
the two
neutral Higgs fields.  The neutralino mass matrix involves the same three
parameters $\mu$, $m_2$, $\tan\beta$, plus $m_1$,
the supersymmetry breaking mass of the $\s b$.

Chargino and neutralino mixing is not an added complication that one may 
introduce into supersymmetric models if one wishes.  It is an intrinsic
feature of these models which must be resolved experimentally. Unless this
can be done, supersymmetry measurements can only be interpreted in  
the context
of model assumptions.  In addition, this measurement is important in
resolving the question of whether the lightest neutralino in supersymmetry
can provide the cosmological dark matter.  In most  scenarios of the dark 
matter, the neutralino must be light and dominantly gaugino rather than 
Higgsino.  In any case, the neutralino mixing must be known to build a 
quantitative theory of the cosmological neutralino production and relic
abundance.

\begin{figure}
\centerline{\epsfxsize=4.00truein \epsfbox{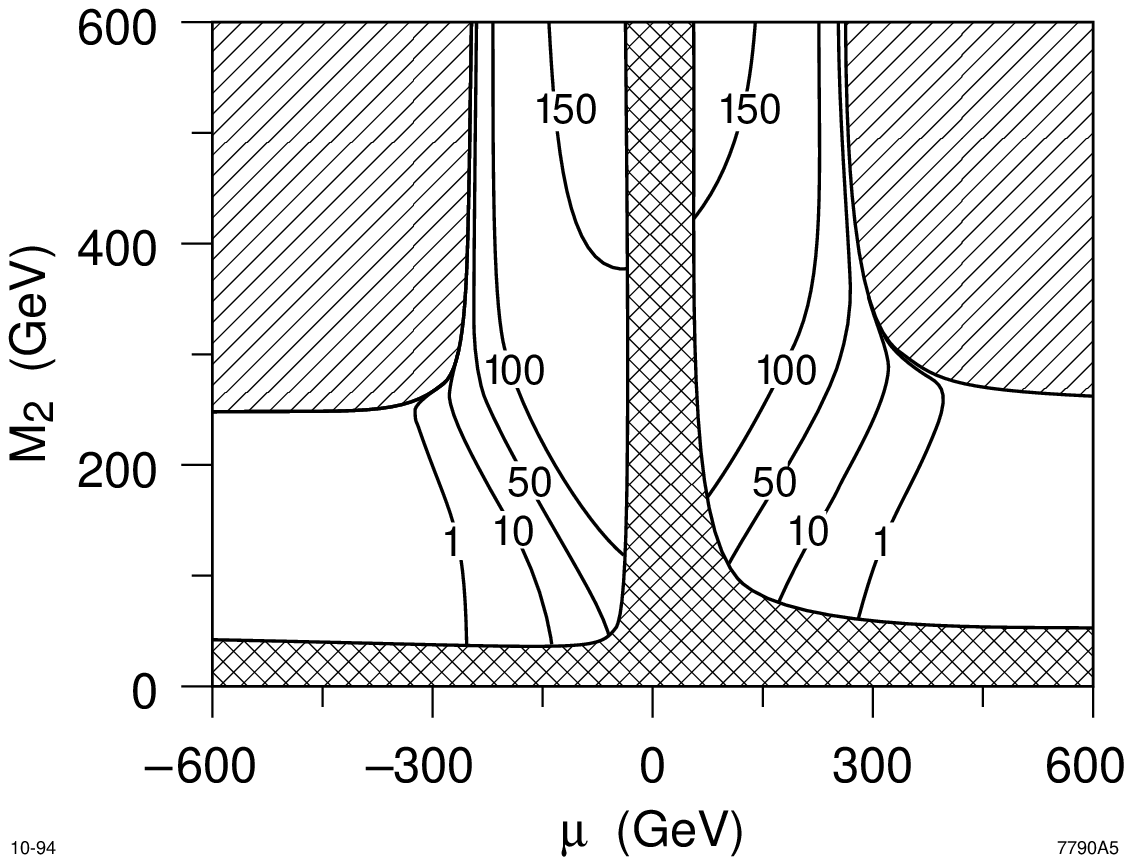}}
 \caption{\small Total cross section
 for $e^-_R e^+ \to \ch1 \chm1$, in fb,  as a 
function of the chargino mass parameters $m_2$ and $\mu$.}
\label{fig:Fengch}
\end{figure}

 Fortunately, it is possible to measure the
chargino and neutralino mixing angles by making use of the
special handles that the linear collider offers.
To see this, consider the diagrams of Fig.~\ref{fig:chdiags}
for a right-handed polarized electron beam.  The second diagram, which
involves the sneutrino, couples only to left-handed electrons and so 
vanishes in this case. At high energy, the $\gamma$ and $Z$ exchanged
in the first diagram can be traded for the neutral $SU(2)$ and $U(1)$ 
gauge bosons.  The $e^-_R$ does not couple to the $SU(2)$ boson.  The 
$\s w^\pm$ does not couple to the $U(1)$ boson.  Thus, the total  
cross section
for the process  $e^-_R e^+ \to \ch1 \chm1$ can  be large only if the 
lighter charginos
$\ch1$ and  $\chm1$ are dominantly composed  of the Higgs field  
superpartners.
This remarkable feature is evident in the contour map of this cross  
section
against $\mu$ and $m_2$
shown in  Fig.~\ref{fig:Fengch}.  A more detailed analysis shows that, by 
measuring the angular distribution of chargino pair production, one can 
determine the separate mixing angles for the positive and negative
(left-handed) charginos \cite{FMPT}.  Unless the mixing angles are  
very small,
the measurement of the two mixing angles and the $\ch 1$ mass allow
the complete mass matrix \leqn{chmatrix} to be reconstructed.  In an
example studied in \cite{FMPT}, this analysis gave a 10\% measurement of 
$\tan\beta$, purely from supersymmetry measurements, in a 100 fb$^{-1}$
experiment at 500 GeV.

Having determined the chargino mixing, one can then analyze chargino
pair production from left-handed fermions.  This brings back the
dependence on the sneutrino mass.  In fact, it is possible to measure
the effect of sneutrino exchange and thus to determine the masses of the 
left-handed sleptons for slepton masses up to a factor of  2 above the  
collider
center of mass energy.   Measurements of the ratio of leptonic to
hadronic chargino decays also can give information on the masses of the 
left-handed sleptons \cite{FS}.  This can provide a consistency test  
on the
supersymmetry parameters or a target for an energy upgrade.

In both the chargino and slepton studies that we have discussed, it is 
remarkable how the use of polarization and detailed angular distribution
measurements can offer new information along a dimension quite  
orthogonal to
that probed by simple mass determinations.  The use of beam  
polarization is
particularly incisive in separating complex composite observables into 
quantities with a direct relation to the parameters in the underlying 
Lagrangian.

\subsection{Studies of the top quark}

The top quark's special status as the most massive known matter particle, 
and the only fermion with an unsuppressed coupling to the agents of
electroweak symmetry breaking, make it a prime target for all future
colliders.  The linear collider, operating near the top quark  
pair-production
threshold and at higher energies below 500 GeV, can carry out a complete
program of top quark physics.  This includes the measurement of the  
top quark
mass, width, form factors, and couplings to many species.
  This broad program of measurements is
reviewed in \cite{FreyGerdes}.  In this section, we will discuss two
particularly important measurements from this collection.

The mass of the top quark is a fundamental parameter in its own  
right, and it
is also an ingredient in precision electroweak analyses and theories
of flavor.  It is important to measure this parameter as accurately as 
possible.  Future measurements at the Tevatron and the LHC are likely to 
determine $m_t$ to 2--3 GeV precision, dominated by systematic
effects \cite{Tev33,ATLAS}.

\begin{figure}
\centerline{\epsfxsize=6.00truein \epsfbox{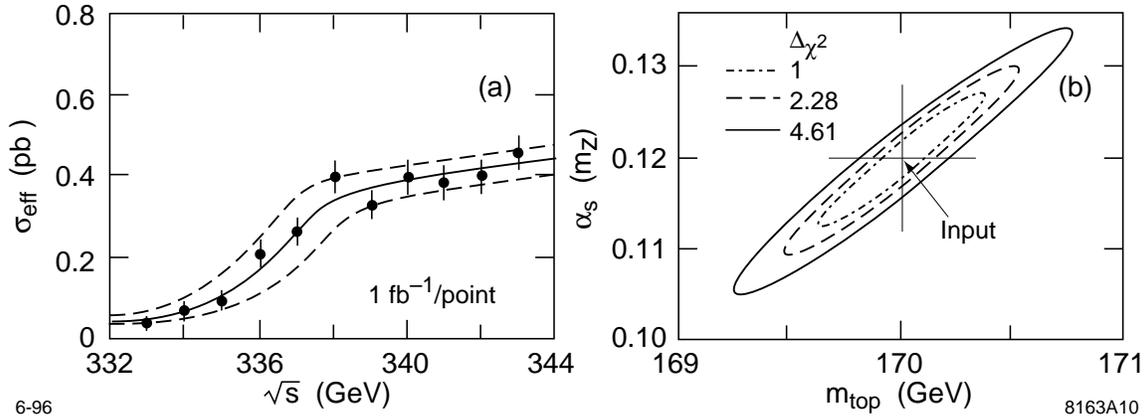}}
 \caption{\small Measurement of the top quark mass from the  
threshold shape,
       using a threshold scan with a total data sample of 11 fb$^{-1}$.
    The effects of beamstrahlung, initial state radiation, and accelerator
      energy spread are included.  A top quark mass of 170 GeV was  
assumed
      in this study \cite{Sumino}.}
\label{fig:topthresh}
\end{figure}

At the linear collider, the top quark mass is determined directly by the 
accelerator energy at which one sees the onset of $t \bar t$ production.
A simulation of the top quark threshold scan, from \cite{Sumino}, is 
shown in Fig.~\ref{fig:topthresh}.  Given a measurement of $\alpha_s$ 
from another source, this scan determines $m_t$ to 200 MeV using only 
 11 fb$^{-1}$ of data.  In the part of the cross section described by 
the top quark threshold, the $t$ and $\bar t$ are separated by a distance
small compared to the QCD scale.  This means that the mass determined
from the threshold scan---as opposed to the `pole mass' determined by the
kinematics of high energy production---is a true short-distance quantity
which is free of nonperturbative effects.  The theoretical error for the 
conversion of the $\ee$ threshold position to the $\bar{MS}$ top  
quark mass
relevant to grand unified theories is about 
300 MeV \cite{Hoang,Hoangandfolks}; for the 
pole mass, it is difficult even to estimate this uncertainty. The
expenditure of 100 fb$^{-1}$ at the $t\bar t$ threshold allows additional
measurements that, for example, determine the top quark width to a
few percent precision \cite{topthresholdone,topthresholdtwo,tthreshthree}.

A second important set of measurements is the study of the top quark  
couplings
to $\gamma$, $Z$, $W$. In the reaction $\ee\to t\bar t$, the final
state can be reconstructed as a 6-jet or 4-jet plus $\ell\nu$ system. 
The $b$ jets should  be identified with an efficiency greater than 80\%.
Both the production through $\gamma$ and $Z$ and the decay by $t\to W^+ b$
are maximally parity violating.  Thus, there are many independent
kinematic variables that can be used to constrain the various possible
production
and decay form factors.   A simulation study using 80\% $e^-$ beam  
polarization
but only 10 fb$^{-1}$ of luminosity at 500 GeV showed that it is possible
to simultaneously constrain the whole set of vector and axial vector 
$\gamma$, $Z$, and $W$ form factors of the top quark with errors in the
range  5--10\%  \cite{FreyGerdes}.  This analysis should improve
further with high-luminosity data samples \cite{Hioki}.  Experiments  
at the
linear collider are sensitive at similar levels to anomalous couplings of 
$t\bar t$ to the gluon \cite{topChrom}.

\begin{figure}
\centerline{\epsfxsize=3.70truein \epsfbox{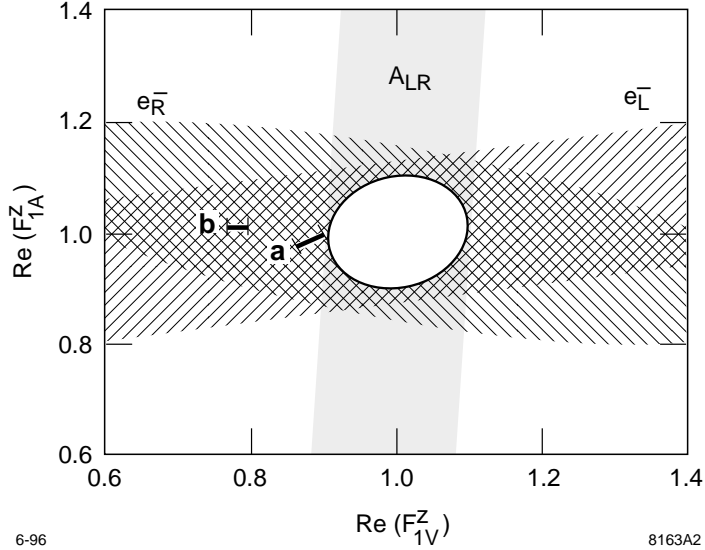}}
 \caption{\small Determination of the form factors for the
  vector and axial vector couplings of the top quark to the $Z$, with 
       100 fb$^{-1}$ at 400 GeV \cite{BarklowTTT}, compared to the
        predictions of technicolor models, from \cite{Murayama}.}
\label{fig:topFs}
\end{figure}

A set of couplings of particular interest are the vector and axial
$t\bar t Z$ form factors.  As we have explained in Section 4.5, these
form factors are predicted to receive large contributions in certain  
models
of strong-interaction electroweak symmetry breaking.  These contributions
result from diagrams in which the
$Z$ couples to the new strongly-interacting species which break  
electroweak
symmetry, and these couple to the top quark through the mechanism which 
generates the top quark mass \cite{CSS}. In Fig.~\ref{fig:topFs},  
the $Z$ form
factor determinations from the simulation study of \cite{BarklowTTT} are
compared to two representative theories \cite{Murayama}.  It is  
interesting
that most of the sensitivity in this particular measurement comes from 
the polarization asymmetry of the total top pair production cross section.
The measurement of this quantity  is dominated by statistics and can be
improved straightforwardly with higher luminosity.

An additional important measurement
 is the determination of the top quark Higgs
Yukawa coupling.   At the LHC,
the ratio $\lambda_{t\bar t h}/\lambda_{WWh}$ can be measured to an
accuracy of 25\% for $80 < m_h < 120$ GeV \cite{ATLAS}.  At a linear  
collider,
the top quark Yukawa coupling can be measured by studying the process
$\ee \to t\bar t h^0$, relying on the $b \bar b$ decay of the $h^0$ to 
produce spectacular events with 4 $b$'s in the final state.  This
process is difficult to study at 500 GeV, but it becomes tractable at
higher energy.  In simulation studies at 800 GeV, where
the cross section is about 8  times higher than at 500 GeV,
 a 1000 fb$^{-1}$ sample yields a 6\% uncertainty on
$\lambda_{t\bar t h}$ for a 120 GeV Higgs boson \cite{BDR,AJuste}.

\subsection{Studies of $W$ boson couplings}

Recent experiments at LEP 2 and the Tevatron have observed weak boson
pair production and have verified the general expectations for the
cross sections given by the Standard Model \cite{wudka,lep2}.  This  
is already
an important discovery.  One of the motivations for building a model  
of the
weak-interaction bosons from  a Yang-Mills gauge theory is that the
special properties of the Yang-Mills coupling tame the typically bad high
energy behavior of massive vector fields.  We now know that the behavior
of the $W$ and $Z$ production cross sections, at least in the region close
to threshold, conforms to the gauge theory predictions.

This discovery sets the stage for the use of $W$ and $Z$ bosons to probe
the physics of electroweak symmetry breaking.  As we have noted in  
Section
4.5, new strong interactions that might be responsible for electroweak 
symmetry breaking can affect the three- and four-particle couplings of the
weak vector bosons.  The precision measurement of
 these effects---and the corresponding
effects on  the top quark couplings discussed in the previous  
section---can
provide a window into the dynamics of electroweak symmetry breaking
complementary to that from direct $W$ boson scattering.

Our discussion  in Section 4.5 implies that a high level of
precision is necessary.  We estimated there that effects of new strong 
interactions affect the standard parameters used to describe the  
$WW\gamma$
and $WWZ$ vertices---~$\kappa_{V}$, $\lambda_{V}$, for $V = \gamma,  
Z$, and
$g_Z$~---at the level of a few parts in $10^{-3}$. For comparison, the
one-loop radiative corrections to these parameters predicted in the
Standard Model are of the order of $10^{-3}$--$10^{-4}$ \cite{argyres}.

In contrast, the
current bounds on  parameters of the $W$ vertices
from LEP 2 and the Tevatron are at the
level of $10^{-1}$ \cite{wudka,lep2,d0run1}.  Much improved  
constraints are
expected from the LHC.  There one expects to place bounds on the $WWV$ 
couplings in the range \cite{ATLAS,lhcwg}
\beq
\left\vert\Delta\kappa_V\right\vert < 0.01\ \mbox{to} \   0.1,
 \qquad \left\vert\Delta
g_1^Z\right\vert,~\left\vert\lambda_V\right\vert < 0.001\ \mbox{to}  
\  0.01
\eeq{LHCvertices}
which might be sensitive to effects of new physics.  It should be noted  
that the LHC analyses integrate over a large range of center-of-mass
energies for vector boson pair production.  This means that the  
sensitivity
and interpretation of these experiments depend on assumptions about the 
energy-dependence of the form factors describing the new physics effects.

The linear collider provides an ideal laboratory for the study of 
the $WWV$ couplings.  The process $\ee\to W^+W^-$ actually gives the  
largest
single contribution to the $\ee$ annihilation cross section at high
energies.  The $W$ pair events can be reconstructed in the  four-jet
final state.  More importantly, the events with a leptonic decay on one
side and a hadronic decay on the other allow unambiguous reconstruction of
the charge and decay angles of the leptonic $W$.
Both the production process and the $W$ decay are strongly
parity-violating, so both beam polarization and angular distributions can 
be used to extract the details of the $W$ vertices.  The
diagrams for $\ee\to W^+W^-$ involve both $\gamma$ and $Z$, but  
these effects
can be disentangled by the use of beam polarization.  The $W$ pair  
production
cross section is about 30 times larger with left-handed than right-handed
polarized beams.  The suppression of the right-handed cross section  
depends
on the relation between the $WW\gamma$ and $WWZ$ vertices predicted  
by the
Standard Model and so is a sensitive measure of deviations from this 
prediction.

    Effects from strong-interaction electroweak symmetry breaking,
which enter through effective Lagrangian parameters as 
in \leqn{forkappa}, affect  
the cross section for longitudinal $W$ pair production through terms
proportional       
to  $(s/m_W^2)$.  At the same time, the fraction of the cross section
with longitudinal $W$ pairs grows as $\beta^2 = (1 - 4 m_W^2/s)$.
From these two effects alone, one should expect a factor 15 improvement
in the sensitivity to these effects in going from LEP 2 to the 
 linear collider experiments at 500 GeV.  The most important advantage,
however, is the 
increase in statistics with high lumnosity running.  A recent simulation
of the $WWV$ coupling measurement at a 500 GeV collider with 500 fb$^{-1}$
estimates the limits that can be placed on the coupling parameters
 as  \cite{burgard}
\beqa
\left\vert\Delta g_1^Z\right\vert < 2.5\times 10^{-3}, \quad
\left\vert\Delta\kappa_Z\right\vert & < 7.9\times 10^{-4}, \quad
\left\vert\lambda_Z\right\vert & < 6.5\times 10^{-4}, \\
\left\vert\Delta\kappa_\gamma\right\vert & < 4.8\times 10^{-4}, \quad
\left\vert\lambda_\gamma\right\vert & < 7.2\times 10^{-4} \ .
\eeqa{lcWlimits}
These results qualitatively improve on the LHC sensitivity, to the point
where not only effects of new physics but even the Standard Model  
radiative
corrections are  visible.

\subsection{Studies of QCD}

In addition to the search for new physics, the linear collider will be 
able to complete the program of precision tests of the Standard Model
with a precise measurement of the QCD coupling constant $\alpha_s$.
The strong coupling constant is determined in $\ee$ annihilation
from the production rate for 3-jet events.
The
reduction in the relative size of hadronization effects at high energy
allow a measurement of $\alpha_s$ with
systematic errors smaller than 1\% \cite{burrowsSnow,Schumm}.

A measurement of $\alpha_s$ of similar quality can be obtained from the 
ratio of hadronic to leptonic decays of the $Z^0$, if one can obtain a 
sample of more than $10^8$ $Z^0$ decays.  This becomes practical in linear
collider experiments at the $Z^0$, as we will explain in Section 5.6.
By comparing the two precision measurements of $\alpha_s$ at $Q$  
values of
$m_Z$ and 500 GeV, it will be possible to give a precise test of the QCD
renormalization group equation.

With confidence in the running of $\alpha_s$ from this experiment,  
one can
extrapolate the precise value of $\alpha_s$ to the grand unification  
scale.
Current data is consistent with a grand unification with the  
renormalization
group equations of supersymmetry; however, it gives little constraint on 
the details of unification. With an accurate  $\alpha_s$, one can  
anticipate
a precise test of grand unification relations.  The contributions to be
accounted for
include next-to-leading order corrections from two-loop beta functions,
TeV-scale threshold effects, and GUT-scale threshold effects \cite{LPol}.
  The two-loop
beta functions are known from the general theoretical scheme.  The  
TeV-scale
threshold effects are unknown today, but they will be determined  
from the new
 particle masses measured at the LHC and the linear collider.  Then a 1\% 
measurement of $\alpha_s$ would allow a 10\% measurement of the GUT-scale
threshold correction.  This measurement would give an indirect but
significant constraint on the spectrum of the massive particles  
responsible for
the GUT level of fundamental symmetry breaking.

The linear collider can also provide the most sensitive experiments
on photon structure,
including the precise measurement of the photon structure function
$F_2^\gamma$.
In addition,
with sufficient forward instrumentation,
the linear collider could study $\gamma^* \gamma^*$ scattering at large
$s$ and fixed momentum transfer.  This is a beautifully clean model system
for analyzing a part of QCD that is still very mysterious, the
nature of the pomeron and the dynamics of high-energy
scattering \cite{BHS}.

\subsection{Precision electroweak studies}

In addition to the experimental program at 500 GeV energies, one can 
envision using the linear collider at the $Z^0$ and the $W$ threshold to 
carry the experimental program of precision electroweak measurements  
to the
next level.
Operation
of the linear collider at the $Z^0$ pole
would yield more than $10^9$ $Z^0$ decays in a
20 fb$^{-1}$ data sample. With more than 100 times LEP 1 statistics and 
high beam
polarization, one could undertake a very ambitious and extensive
program of precision
measurements. For example \cite{Monig,Erler},
employing the left-right polarization
asymmetry, leptonic forward-backward asymmetries, and tau polarization
asymmetry (all of which are currently statistics limited) one could
improve the determination of $\sin^2\theta^{\rm eff}_W$ at the $Z$ pole
by an order of magnitude, bringing it to an unprecedented $\pm 0.01\%$
level.
Other quantities such as the $Z$ line shape parameters,
$R_b$ = $\Gamma(Z\to b\bar b)/\Gamma (Z\to {\rm hadrons})$, and $A_b$ 
(the polarized $b\bar b$ asymmetry) could also be improved. They would be
limited only by systematics.

With such a large sample of $Z$ decays, one would have  more than 
$10^8$ $b\bar b$
and $3\times10^7$ $\tau^+\tau^-$ pairs. The study of these events could 
make use of the outstanding  vertex resolution
and detection
efficiency of the linear collider environment.
In addition, polarized
$\ee$ annihilation at the $Z^0$ produces (for a left-handed beam)
dominantly forward production of $b$ quarks and backward production of 
antiquarks, thus eliminating the need for a flavor tag.
These features combine to give an
ideal environment for studying CP
violating asymmetries and rare decays as well as performing precision
measurements \cite{Monig}.
For example, one could improve the current precision
on the forward-backward asymmetry parameter $A_b$
by more than an order of magnitude.

\begin{table}[t]
\begin{center}
\begin{tabular}{|l|c|c|}\hline
Parameter & Current Value & LC Measurement \\[0.5ex] \hline\hline
$\sin^2\theta^{\rm eff}_W$ & $0.23119\pm0.00021$ & $\pm 0.00002$ \\
$m_W$ & $80.419\pm0.038$ GeV & $\pm0.006$ GeV \\
$\Gamma(Z\to\ell^+\ell^-)$ & $83.96\pm0.09$ MeV & $\pm0.04$ MeV \\
$R^{\rm exp}_b / R^{\rm th}_b$ & $1.0029\pm0.0035$ & $\pm0.0007$ \\
$A^{\rm exp}_b / A^{\rm th}_b$ & $0.958\pm0.017$ & $\pm0.001$ \\[.5ex]
\hline
\end{tabular}
\end{center}
\caption{\small Current values of some important electroweak  
parameters,  and
the potential uncertainty obtainable at a linear collider providing
 with high statistics (\eg, $10^9$ $Z^0$ decays).}
\label{tabone}
\end{table}

In Table~\ref{tabone}, we have listed some improved measurements
envisioned at the linear collider.  The tiny error on
$\sin^2\theta^{\rm eff}_W$ assumes a precise beam polarization  
measurement
that may require polarizing both the electron and positron beams.
 The importance of refining
$\sin^2\theta^{\rm eff}_W$ is well illustrated by the
prediction for the
Higgs mass that would be obtained by employing these precise values
and the improved value of $m_t$ from Section 5.4 as input.  One finds
\beq
m_h = (140\pm 5{\rm~GeV}) e^{[1911 (\sin^2\theta^{\rm eff}_W -
0.23158)]} \ ,
\eeq{higgsM}
where the dominant error comes from hadronic loop uncertainties
in $\alpha$ (assumed here to be reduced by a factor of 3 compared to the
current error). Comparison of the indirect loop determination of $m_h$
from \leqn{higgsM} with the direct measurement of $m_h$ from the LHC and 
the linear collider would confront the electroweak prediction at the 
5\% level and would provide an accurate sum rule to be satisfied by new
heavy particles with electroweak charge.  Another way to look at this
comparison is that it will probe the $S$
and $T$ parameters to an accuracy of $0.02$, about 8 times better
than current constraints. At that level, even the existence of a single
heavy chiral fermion doublet (much less an entire dynamical symmetry
breaking scenario) would manifest itself.  The accurate value of
$\sin^2\theta^{\rm eff}_W$ at the $Z$ pole would be a valuable input 
to the measurements of cross sections and asymmetries at high energy that
we will discuss in Sections 6.3 and 6.4, measurements which probe for 
possible $Z'$ bosons, lepton compositeness, or new space dimensions.

A linear collider run near the the $W^+W^-$ threshold would
also be extremely valuable for improving the determination of $m_W$
beyond the capabilities of the LHC \cite{Erler}.
Already at the current uncertainty
of 40 MeV, the determination of the $m_W$ mass from kinematic fitting
of $W$ pair production
at LEP 2 is affected by
systematic uncertainty from the modeling of fragmentation.  But the
interpretation of the
measurement of the $W$ threshold position is almost free of theoretical
uncertainty, allowing a 6 MeV measurement to be done with a dedicated
100 fb$^{-1}$ run.

Collectively, the broad program of precision electroweak studies which 
the high luminosity of the linear collider makes available
nicely complements and expands the  physics goals at the maximum collider 
energy.

\section{Further topics from the linear collider physics program}

In the preceding section, we have discussed only those aspects of  
the linear
collider experimental
program for which there are strong arguments that the phenomena to  
be studied
will appear at 500 GeV.  There are many other experiments that can be done
at an $\ee$ linear collider which has sufficient energy to reach the 
required threshold for new particles.  In this section, we will describe a
number of experiments of this character.  All of these experiments will
eventually become relevant as components of the long-term program
that we have described in Section 2.
Measurements at the LHC which estimate the new thresholds could provide 
specific  motivation for upgrading a 500 GeV collider to higher energy.
But, one should keep in mind  that all of the phenomena we describe  
in this
section could well be present at 500 GeV and provide additional richness
to the initial physics program of the linear collider.

It is well appreciated that an $\ee$ collider provides an excellent
 environment to search for
all varieties of exotic particles with nonzero electroweak quantum  
numbers.
The huge variety of particles which have been searched for at LEP is 
described, for example,
in \cite{Ruhlmann}.
In almost all cases, the LEP limits are close to the kinematic limit  
allowed
by the collider.
A collider  operating above the pair production threshold will be able to 
accumulate a large sample of events  (70,000 events  per unit of $R$ in 
a 200 fb$^{-1}$ sample at 500 GeV)  and
make incisive measurements.

The corresponding discovery reach for exotic particles at the LHC ranges 
from a few hundred
GeV for new leptons to about 2 TeV for new quarks.  So, as a general 
statement, the locations
of the new thresholds are likely to be found at the LHC.
Experimenters at a linear collider will measure essential
information that is beyond the capability of the LHC.
We have seen examples of this in Section 5, and further
examples will be discussed in this section.

Rather than summarize all possible measurements of new phenomena at
a linear collider, we restrict
ourselves in this section to four specific examples that have been worked 
out in some detail.  In Section 6.1, we will discuss the particles of 
an extended Higgs sector such as that in the Minimal Supersymmetric
Standard Model.  In Section 6.2, we will discuss studies of  
supersymmetric
particles beyond the lightest chargino, neutralinos, and sleptons.
In Section 6.3,
we will discuss  new and exotic $Z^\prime$ bosons.   In Section 6.4, 
we will discuss probes of large extra dimensions
and TeV-scale quantum gravity.

Because this paper focuses on  the issue of a 500 GeV  collider,
we do not discuss here the significant capabilities of higher energy
$\ee$ collisions to
probe $WW$ scattering processes \cite{BoosWW}. These include the unique
ability to study the
reaction $W^+W^-\to t\bar t$, which directly tests the coupling of the 
top quark to the particles responsible for strong-interaction electroweak
symmetry breaking.  These experiments, and the comparison to the LHC
capabilities, are reviewed in \cite{SnowmassSC,HanWWtt}.

Although the detailed physics justification for increased
$e^+e^-$ collision energy is more difficult to quantify at present
than that for the initial $~500$ GeV step, we fully expect
that the experimentation at the LHC and first stage $e^+e^-$
linear collider will reveal phenemena that dictate energy
upgrades.  It is important to continue the R\&D needed for this
evolution.

\subsection{Extended Higgs sector}

In Section 5.1, we have discussed the measurement of the properties  
of the
lightest Higgs boson.  Many models of new physics allow multiple Higgs 
fields, leading to additional heavier Higgs particles.  In particular, 
supersymmetry requires the presence of at least two Higgs doublet  
fields.
This produces, in addition to the $h^0$, four additional states---the  
CP-even $H^0$,
the CP-odd $A^0$, and charged states $H^\pm$.   The masses of these  
states
should be comparable to the masses of other supersymmetric particles.  If 
the scale of superparticle masses is much greater than 100 GeV, then
 typically the four heavy Higgs states are relatively close in mass, 
and the light $h^0$ resembles the Higgs boson of the Standard Model.

The heavy Higgs states are very difficult to find at the LHC.  The LHC 
experiments have studied extensively their sensitivity to the Higgs
sector of the MSSM.  We have already presented a summary of these
analyses in Fig.~\ref{fig:LHCHiggs}.
A low mass $H^\pm$ can be found at the LHC below about 125 GeV in  
the decays
of the top quark.  For $m_{H^\pm}$ above 225 GeV, its decay into
$t \bar b$ can be used to find the charged Higgs if $\tan\beta\gsim 25$ or
$\tan\beta \lsim 2$.
 In the region of intermediate
$\tan\beta$ above the LEP limits, only the process $h^0 \to \gamma\gamma$ 
is visible, and the $H$ and $A$ are not seen at all.
  For larger $\tan\beta$ ($>10$), the decays $H/A
\rightarrow \tau^+\tau^-$ become accessible.  Because the  technique for
detecting $H$ and $A$ involves particles that decay with missing energy,
it will be difficult to make a precise mass measurement.
ATLAS studies suggest an accuracy on the $H$/$A$ mass of about 5 GeV, 
 for $M_{H/A}=300$ GeV
and $\tan\beta=10$,  only
after 300 fb$^{-1}$ has been collected. 
For comparison, the $H$--$A$ mass difference is
at most a few GeV.  For low $\tan \beta$, $H$ could be
detected by $H\rightarrow ZZ^*$.
This mode, however, applies only to a limited region of parameter
space, $\tan \beta < 3$ (a region disfavored by the LEP constraint on the 
mass of $h$) and $m_H < 350$ GeV.

A crucial aspect of the experimental study of the heavy Higgs states  
would be
to measure the value of $\tan\beta = \VEV{\phi_2}/\VEV{\phi_1}$,
where $\phi_1$ and $\phi_2$ are the two Higgs doublets
of the MSSM.    This
quantity is needed to determine the absolute size of the quark and lepton 
Yukawa couplings.  For example, it is possible that the bottom quark  
Yukawa
coupling is large and the lightness of the bottom quark is explained  
by the
fact that the Higgs field responsible for this mass has a small vacuum 
expectation  value.  In supersymmetry, $\tan\beta$ also appears in many
formulae for the supersymmetry masses and mixings and is a source of 
theoretical uncertainty unless it can be pinned down.  The LHC can measure
$\tan \beta$ from the heavy Higgs particles only where $H$ is visible by 
one of the techniques just listed, to an accuracy of 10--30\%.  It should
be noted that what is measured is $\sigma\cdot BR$, and so the  
determination of
$\tan\beta$ depends on theoretical assumptions about the total width.

If the masses of $H$, $A$ are well above that of $h$, these particles are 
mainly produced at an $\ee$ collider in pairs, through $\ee\to H^0  
A^0$.  The
mass determination is straightforward.  Kinematic fitting of decays with 
$b \bar b$ on both sides should give an accuracy of 0.3\%.  The
program described earlier for the precision determination of the $h$
branching fractions can be applied also to the $H$ and $A$.  The crucial 
parameter $\tan \beta$ is given by the ratio of the branching ratios to 
$b\bar b$ and $t \bar t$.  For $A$,
\beq
 {\Gamma(A\to t\bar t) \over \Gamma (A\to b\bar b) } = {m_t^2\over m_b^2}
      \cot^4\beta \cdot \left(1 - {4m_t^2\over m_A^2}\right)^{1/2} \ .       
\eeq{btHABRs}
From this measurement, a completely model-independent
determination of $\tan\beta$ to 10\% accuracy is expected.
Measurements of other branching fractions of $H$, $A$, and $H^\pm$ will
provide cross-checks of this value \cite{FengMoroi}.

The ATLAS \cite{ATLAS} and CMS \cite{CMS} 
analyses of the fitting of LHC data to the minimal
supergravity-mediated model gives a remarkable accuracy of 3\% in the 
determination of $\tan\beta$.  
However, this determination of $\tan\beta$
is based on the assumption of a specific model of supersymmetry breaking.
It uses the  precision measurement of the $h^0$ mass and thus depends  
on the
detailed theory of the one-loop supersymmetry corrections to this parameter. 
Linear collider experiments offer a number of methods to determine
$\tan\beta$ from supersymmetry  observables in a model-independent way.
For example, $\tan\beta$ can be extracted from chargino mixing, as we
have discussed in Section 5.2.4.  In the end, it is a nontrivial test of the 
theory whether the determinations of $\tan\beta$ from the 
supersymmetry  spectrum
agree with the direct determination of this parameter from the Higgs  
sector.

\subsection{Supersymmetric particle studies}

In Section 4.4, we have argued that, if the new physics at the TeV
scale includes supersymmetry, the lightest supersymmetric particles are
likely to appear at a 500 GeV $\ee$ collider.  In Section 5.2, we have 
discussed the program of detailed measurements on those particles.  Of 
course, nothing precludes a larger set of supersymmetric particles from
appearing at 500 GeV, though it is likely that  increased energy
will be needed to produce the full supersymmetry spectrum.
In this
section, we will discuss what can be learned from a more complete
study of the supersymmetry spectrum in $\ee$ annihilation.

For brevity, we focus on two important issues.  The first of these
is whether  supersymmetry does in fact give
 the dynamics that leads to electroweak
symmetry breaking.  To verify the mechanism of electroweak symmetry
breaking experimentally, we must determine the basic parameters that
directly determine the Higgs potential. These include the heavy Higgs
boson masses discussed in the previous section.  Another essential
parameter is $\mu$,  the
supersymmetric Higgs mass parameter.  As we have discussed in Section 5.2,
this parameter can already be determined from the study of the lighter
chargino if these particles are not almost pure $\s w$.  In that  
last case,
$\mu$ is determined by measuring the mass of the heavier charginos.
We have argued in Section 4.4 that these particles should be found
with at most a modest step in energy above 500 GeV.
A precision mass measurement can be done using the endpoint technique 
discussed in Section 5.2.

In typical supersymmetric models, the negative Higgs (mass)$^2$ which
causes electroweak symmetry breaking is due to a mass renormalization 
involving the top squarks.  This same renormalization leads to
${\s t}_L$--${\s t}_R$ mixing and to a downward shift in the top squark
masses relative to the masses of the first- and second-generation squarks.
The mass shift, at least, might be measured at the LHC.  However, in some
scenarios with a large mass shift, only the third-generation squark
masses can be measured accurately \cite{ATLAS}.  At the linear collider, 
flavor-dependent squark masses can be measured to accuracies better than 
1\%.  In addition, the mass differences of the partners of $q_L$  
and $q_R$
can be measured to this accuracy using polarization asymmetries  
\cite{FFin}.
By comparing the pair production cross sections with polarized beams,
as described in Section 5.2 for stau mixing, it is possible to measure
the top squark mixing angle to better than
1\% accuracy in a 500 fb$^{-1}$ experiment \cite{Bartl}.

\begin{figure}
\centerline{\epsfxsize=6.00truein \epsfbox{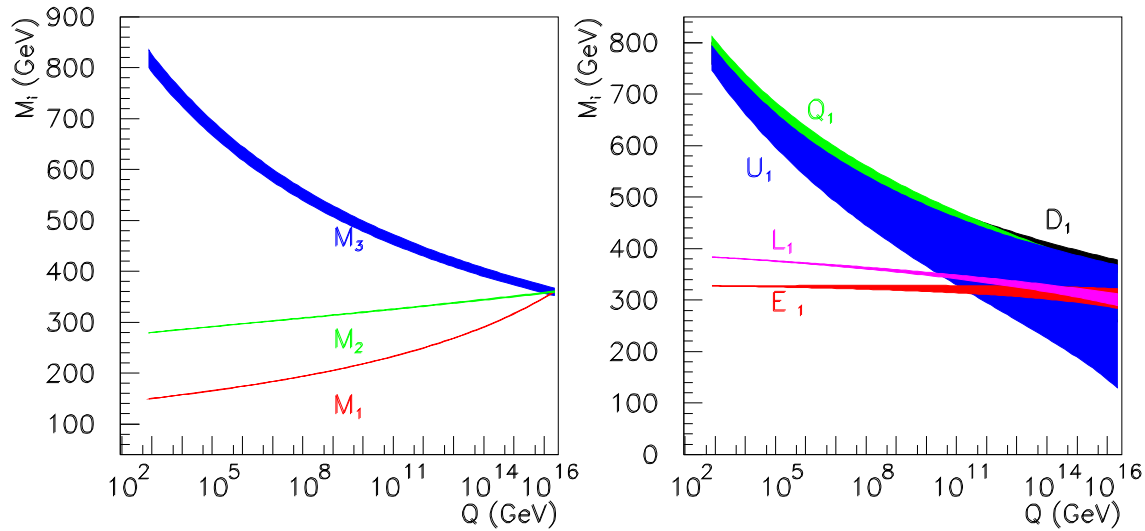}}
 \caption{\small Extrapolation of supersymmetry mass parameters  
determined
at a linear collider from the TeV scale to the grand unification scale,
     from \cite{Zerwas}.  The width of each band at the weak scale is the
      error in the direct parameter determination; these errors are
     propagated to higher energies using the renormalization group  
equations.}
\label{fig:Zerwas}
\end{figure}

The second issue is the possibility of the grand unification
of supersymmetry breaking parameters.  This is the crucial test of
whether supersymmetry breaking arises from physics above the grand
unification scale or from a different mechanism acting at lower energies.
This test requires accurate model-independent determinations of as many
supersymmetry mass parameters as possible.  Figure~\ref{fig:Zerwas} shows
an extrapolation to the grand unification scale
at $2\times 10^{16}$ GeV of masses determined
in a 500 fb$^{-1}$ sample at a linear collider.  The most effective
tests of grand unification come from the comparison of the gaugino
mass parameters $m_1$ and $m_2$ and from comparison of the  masses of 
the sleptons  ${\s e}_R$ and ${\s e}_L$ (called $E_1$ and $L_1$ in the
figure).  Because of QCD
threshold corrections, the masses of the gluino ($m_3$) and
the first-generation
squarks
(labeled $D_1$, $Q_1$, $U_1$) are
less effective in this comparison.  It should be noted that the mass  
ratios
which provide the most significant tests of grand unification
 are just the ones that are
most difficult to measure accurately at the LHC.
Even for the uncolored states, a 1\%
mass error at the weak scale evolves to a 10\% uncertainty at the grand 
unification scale.  So this comparison puts a premium on very precise 
mass determinations, such as a linear collider will make possible.

These issues are only two slices through the rich phenomenology of
supersymmetric particles.  If supersymmetric particles---or any other
family of exotic particles---appear at the TeV scale, there will be
a full program of experiments for both hadron and $\ee$ colliders.

\subsection{New $Z'$ bosons}

The new physics at the TeV scale must have $SU(3)\times SU(2)\times U(1)$
gauge symmetry, but it might have an even larger gauge symmetry with  
additional
heavy vector particles.   The simplest extensions are those with
extra $U(1)$ gauge symmetries.  The corresponding gauge bosons appear as 
new vector resonances---$Z'$ bosons---coupling to lepton and to  
$q\bar q$
pairs.

Extra $U(1)$ factors in the gauge group preserve the predictions
of grand unification.  In fact, these new symmetries appear naturally in 
models in which the grand unification group is larger than the minimal
choice of $SU(5)$.  For example, the grand unification group $E_6$
contains the Standard Model gauge group and two additional $U(1)$ factors.
This leads to models in which the gauge symmetry at TeV energies
contains an additional $U(1)$ factor which is a linear combination of 
these \cite{HewettPR,Leike}.

In certain grand unified models, the masses of the heavy neutral  
leptons which
give the scale of the neutrino mass seesaw are determined by the scale of 
breaking of an extra $U(1)$ symmetry.  In this case, the extreme lightness
of neutrinos puts the mass of the $Z'$ beyond the reach of accelerator
experiments.  But many other motivations for a new $U(1)$ symmetry point
to lower masses \cite{CveticLsusy}.
In particular, the size of the $\mu$ parameter of supersymmetry
may be controlled by the scale of breaking of a $U(1)$ symmetry,
in which case the corresponding $Z'$ boson must have a mass not
far above 1 TeV.  More generally, the possible richness of gauge
symmetries motivates the search for these new states.  This is
especially true for  superstring theories, where explicit model
constructions often  predict a large number
of extra $U(1)$ gauge particles \cite{CveticLStr}.

The abilities of colliders to detect signatures of heavy $Z'$ bosons have 
been studied in great detail.  Hadron colliders have impressive  
sensitivity
for searches in which the $Z'$ bosons appear as resonances decaying to 
$\ell^+\ell^-$.  Lepton colliders can be sensitive to $Z'$ bosons in 
a different way, through the precision study of the pair production processes
$\ee\to \ell^+\ell^-$ and $\ee \to q\bar q$.  Because these  
reactions can be
measured precisely and also predicted theoretically to part per mil  
accuracy,
experiments can be sensitive to interference effects caused by $Z'$ bosons
of mass a factor of 10 or more above the $\ee$ center of mass
energy \cite{LeikeR,CveticG,DelAguila}.  All of the special handles of the
$\ee$ environment, including polarization asymmetries, flavor tagging, and
$\tau$ polarization, can be brought to bear in the search for these
interference effects.

Table~5, based on \cite{Rizzo}, gives a comparison
between the sensitivity of $\ee$ linear colliders and that of the
LHC.  The models listed in the table correspond to particular choices for 
the quantum number assignments of the $Z'$; see the original  
reference for
details.  The table shows that the sensitivity of a linear collider
operating at 500 GeV is quite comparable
to that of the LHC.  The sensitivities quoted in the table correspond to 
different types of measurements, and this point illustrates the 
complementary relation of the LHC and the linear collider.
  For a $Z'$ at
a few TeV, the LHC will identify a resonance and accurately measure the
mass $M$.  The linear collider will measure interference effects and thus
determine the quantity  $g_e g_f/M^2$ which depends on
 the mass and the coupling strengths to the electron and the flavor $f$.
By combining these pieces of information, one may obtain a complete
phenomenological profile of the $Z'$.  Both hadron and lepton collider
experiments will thus be needed to understand
how the $Z'$ fits into the larger picture of unification and symmetry.

\begin{table}
\begin{center}
\begin{tabular}{|l|c|c|c|}\hline
  Model   &  500 GeV    &   1000 GeV &  LHC \\[.5ex] \hline\hline
$\chi$  &  4.5   &   6.5   &  4.5  \\
$\psi$  &  2.6  &    3.8  & 4.1\\
$\eta$  &  3.3 &    4.7  & 4.2\\
I       &  4.5 &    6.5 & 4.4\\
SSM     &  5.6 &    8.1 & 4.9\\
ALRM    &  5.4 &    7.9 & 5.2\\
LRM     &  5.2 &    7.5 & 4.5\\
UUM     &  6.7 &    9.8 & 4.6 \\[.5ex] \hline
\end{tabular}
\caption{\small Sensitivity of $\ee$ linear colliders and the LHC to  
effects of
a $Z'$, after \cite{Rizzo}. The table gives the mass reach in TeV for
observability at the 95\% CL.
 The analysis for linear colliders is based on
measurement of indirect effects for an event sample of 200 fb$^{-1}$; it 
includes the effect of experimental cuts.  The analysis for the LHC
gives the direct sensitivity to a resonance, assuming an event sample of 
100 fb$^{-1}$ and $Z'$ decays only to Standard Model fermions.}
\end{center}
\label{tab:Rizzo}
\end{table}

This study of $\ee\to f\bar f$ can also be used to search for composite
structure of quarks and leptons.  The process most sensitive
 to compositeness is Bhabha scattering.
  A 200 fb$^{-1}$ experiment at 500 GeV would be
expected to place a limit of 90 TeV on the $\Lambda$ parameters of
electron compositeness.  M\o ller scattering ($e^-e^-\to e^-e^-$)
potentially provides an even more
sensitive probe, offering a limit of 130 TeV for a 200 fb$^{-1}$  
experiment
at 500 GeV \cite{BarklowComp}.  Even the $\ee$ limit is a factor of 6
above the expected limit from studies of Drell-Yan production at
the LHC \cite{ATLAS}. In addition, an effect seen at the LHC could come
from any one of a large number of possible operators, while in polarized
Bhabha or M\o ller scattering the operator structure can be determined
uniquely.

\subsection{Large extra dimensions}

Among the most remarkable proposals for new physics at the TeV scale is
the idea that new space dimensions play an important role.  String
theorists have insisted for many years that Nature contains more
than four dimensions.  However, for a long time the extra dimensions
were considered to be unobservably small.  Recently, new developments
in string theory and phenomenology have shaken up this complacent
picture and have suggested that new space dimensions may be of the
size $\hbar$/TeV, or even larger \cite{Antoniadis,Lykken,AHAD}.

There is no space here for a complete review of these new developments.
(A brief review can be found in \cite{Tampere}.)  But we would like to 
indicate the role that the LHC and the linear collider could play in the
elucidation of these models.

Consider first models in which there is a single new dimension of  
TeV size.
In this model, the basic quantum fields in Nature are five-dimensional.
The momentum in the fifth dimension is quantized and can be  
interpreted as
the mass of a four-dimensional field.  So, each quantized value of the 
fifth component of momentum gives a state that we would observe as a 
new heavy particle. The easiest states to observe are the components  
of the
photon and $Z$ with nonzero momentum in the fifth dimension.  These
would appear as $Z'$ bosons.  The sensitivity of the LHC and the linear
collider to these states is greater than that to the `SSM' (Sequential
Standard Model) boson listed in Table~5.  If several states
can be discovered, one can begin to map out the geometry of
the extra dimensions.  A similar phenenomenology applies to the
Randall-Sundrum model \cite{RSed}
in which curvature in the fifth dimension is used
to explain the hierarchy between the Planck scale and the weak scale.
In this case, the new resonances are actually higher Fourier components of
the gravitational field, a fact which can be recognized experimentally 
by their characteristic spin-2 decay distributions \cite{DHR}.

In another class of models, our apparently four-dimensional world is a
membrane in a space of larger dimensionality \cite{AHAD}. This  
scheme allows
the scale at which quantum gravity becomes a strong interaction to be 
much lower than the apparent Planck scale.  In fact, it can be as low as 
TeV energies.
The authors of \cite{AHAD} emphasized that their theory could be
 tested by macroscopic gravity experiments. But in fact
more stringent tests come from high energy physics, from experiments
that look for the effects of gravitational radiation at high energy
colliders.  These are of two types.  First, if the scale $M$ of
strong quantum gravity
is low, one expects radiation of gravitons $G$ in $\ee$ and $q\bar q$ 
collisions, giving rise to processes such as
\beq
     \ee \to \gamma G \, \qquad q \bar q \to g G \
\eeq{missingG}
which appear as photons or jets recoiling against an unobserved
particle.
These effects have been searched for explicitly at LEP and the Tevatron
(\eg, \cite{AlephG}), giving lower limits of about 1 TeV on the gravity
scale $M$.  Second, one can look for the effects of virtual graviton 
exchange interfering with Standard Model annihilation processes.
These interference effects have been  searched for
 both by  measurements of $\ee$ annihilation
to fermion pairs at LEP 2 (\eg, \cite{Lthree}) and by measurements of 
Drell-Yan and $\gamma\gamma$ pair production at the Tevatron  
\cite{DzeroGG}.
In both cases, the sensitivity to $M$ reaches above 1 TeV.

These experiments will be repeated at the next generation of colliders.
The limits on $M$ from
missing energy experiments are expected to be about 5 TeV from the
high luminosity linear collider at 500 GeV, and about 8 TeV from monojet 
searches at the LHC.  Similarly, limits on $M$ from virtual graviton
exchange should reach to about 6 TeV both at the 500 GeV linear collider
and in the study of Drell-Yan processes at the LHC \cite{HewettSit}.
These values are high enough that, if the new dimensions are actually 
connected to the physics of the TeV scale, their effects should be  
observed.
In that case, the linear collider experiments will take on an added
significance.  At the linear collider, but not at the LHC, it is possible
to determine the parton kinematics of a missing energy event.  Then one
can determine whether events have a broad mass spectrum, as predicted in 
ordinary quantum gravity, or whether they are resonant at fixed mass 
values, as predicted in string theory.  For virtual graviton processes, 
the linear collider can observe the flavor- and helicity-dependence of the
interference effects and determine whether the new couplings are  
universal,
as naively expected for gravity, or are more complex in nature.

If there are more than four dimensions in Nature, the evidence for this 
will most likely
come from high-energy physics.  The possibility provides a
tremendous opportunity, one which will engage experimenters at both
hadron and lepton colliders.

\section{Conclusions}

The beautiful experiments in particle physics over the past
twenty years and the tremendous theoretical effort to
synthesize the current understanding of electroweak symmetry
breaking have brought us to a point of exceptional opportunity
for uncovering new laws of physics.  The wealth of precision
electroweak measurements indicate that a new threshold is
close at hand.   The precision measurements place strong
constraints on models that explain the symmetry breaking
and point to new phenomena at the 500 GeV scale.

Later in this decade, we will begin to capitalize on this
opportunity with experiments at the LHC.  There is no doubt
that the LHC will make important discoveries.  However, many
crucial measurements on the expected new physics are difficult
to perform at a hadron collider.  In this paper we have
argued that a 500 GeV linear collider will provide essential
information needed to interpret and to exploit these
discoveries.

The LHC should discover a Higgs boson (if LEP 2 or
Tevatron experiments have not already done so) in all but
rather special circumstances.  The linear collider is
very well suited to measuring its quantum numbers, total
width and couplings.  Moreover, if there is an expanded
Higgs sector, measurement of the Higgs couplings to fermion
pairs and to gauge bosons is essential.

If the new physics includes supersymmetry, the LHC experiments
should observe supersymmetric particle production.  They will
measure some fraction of the sparticle masses, but they most
likely will not be able to determine their spin and electroweak
quantum numbers.
Measurement of mixing angles and supersymmetric couplings at
the LHC will be very difficult.
To the extent that the sparticles
are accessible to a linear collider, these measurements
are straightforward and precise.  We have argued that there is
a good probability that some of the crucial sparticles will be
within reach of a 500 GeV collider.   The measurements of gaugino
and sfermion mixings and masses will provide important clues
towards understanding how supersymmetry is broken and
transmitted to the TeV scale.

We have reviewed the models in which new strong interactions
provide the means by which the Standard Model particles acquire
mass, and have found that although such models cannot be ruled
out, they have become increasingly constrained by the existing
precision data.   The LHC has the possibility for
observing new strong interactions through modifications to
$WW$ scattering.   We have argued that analogous modifications
to the gauge boson or top quark couplings can be seen with a 500
GeV  linear collider.  We have also suggested that operation of
the linear collider at the $Z$ resonance may be profitable.

In each of these examples, we have argued that the linear collider
and the LHC have complementary roles to play.  It is likely that
neither machine, by itself, will piece together the full picture
of electroweak symmetry breaking.  The strength of the LHC is its
large partonic energy and copious production of many new particles.
The linear collider, with its control of partonic energy and beam
polarization, and with favorable signal to background ratios, can
make crucial measurements that reveal the character of new
phenomena.   The complementarity of hadron and lepton collisions
has been amply demonstrated in the past, and there is every reason
to expect that it will continue in the future.

It may be useful to give a few illustrative examples of how
the linear collider program might respond to possible outcomes
of the LHC experiments:
\begin{enumerate}
\item {\em A Higgs-like state is discovered below 150 GeV, and
strong evidence for supersymmetry is found.}  In this case, the
linear collider program would be based primarily on the exploration
of supersymmetry and the extended Higgs sector.  It would measure
the couplings, quantum numbers, mixing angles and CP properties
of the new states.  These precisely measured parameters hold the
key for understanding the mechanism of supersymmetry breaking.
In this scenario, a premium would be placed on running at sufficiently
high energy that the sparticles are produced.   This might dictate
raising the energy to at least 1 TeV.
\item {\em A Higgs particle is seen, and no evidence for
supersymmetry is found.}
The key objective in this scenario would be the thorough
investigation of the Higgs particle.  Here, precision
measurements would be of paramount importance; a linear
collider would be able to make precise determinations of the
Higgs couplings to all particles (including invisible states),
as well as of its total width, quantum numbers and
perhaps even the strength of its self coupling.   Such
measurements would point the way to possible extensions of
the Standard Model.

High luminosity operation would be necessary at the optimum
energy for Higgs production.  In this scenario, revisiting
the $Z$ pole might be critical to refine knowledge
of electroweak loop corrections.  Increased energy would
likely be required to search for new phenomena
such as strong scattering of $WW$ pairs or evidence for
large extra dimensions.
\item {\em No new particles are found.}
This uncomfortable scenario extends the puzzlement we are
in today.  In this case the first goal of a linear collider
would be to close the loopholes in the LHC measurements
(such as the possibility that the Higgs decays dominantly
to invisible particles).  After that, a detailed study of
the top quark or gauge boson couplings would be necessary
to reveal evidence
for new dynamics.  In this scenario, increased energy
would be necessary to study $WW$ scattering.  One might
wish to carry out additional precise
measurements at the $Z^0$ pole.
\item {\em A wealth of new phenomena is sighted at LEP,
Tevatron and LHC.}
These discoveries would indicate a much richer array of new
particles and phenomena than are presently envisioned in any
single model.  In this case, with multiple sources of new
physics, the job of the linear collider is clear.  With
its unparalleled ability to make detailed measurements of
the properties of the new states, a linear collider would be
essential to map out the terrain. A long and rich program
would be assured.
\end{enumerate}

In each of these representative scenarios, after examination
of the many ways that new physics might come into view, we
conclude that a linear collider has a decisive role to play.
Starting with initial operation at 500 GeV, and continuing
to higher energies as needed, an $\ee$ linear collider would
be at the heart of a rich twenty-year program of experimentation
and discovery in high energy physics.

There is no guarantee in physics that we can ever predict how
Nature chooses to operate in uncharted territory.   Over the
past two decades, however, through theory and experiment, a
remarkable understanding has developed.  In this paper we
have argued that the data offer a clear picture of how
the next step should proceed:  We should begin the detailed design
and construction of a 500 GeV $\ee$ linear collider.

\Acknowledgements

We are grateful to many colleagues in the US, Canada, Europe, and Japan
for the insights into linear collider physics which are reflected in this
document.  This work was supported by grants to the authors from the 
 US Department of Energy and the US National  
Science
Foundation.



\end{document}